\documentclass[useAMS,usenatbib,usegraphicx]{mn2e}

\usepackage{amssymb}

\newcommand{\eqref}[1]{\ref{#1}}

\title[Topology in the 2dFGRS]{Topology of large-scale structure in the 2dF Galaxy Redshift Survey}
\author[James, Lewis and Colless]{J. Berian James$^{1,2}$\thanks{E-mail:
jbjames,gfl@physics.usyd.edu.au; colless@aao.gov.au}, Geraint F. Lewis$^{1}$ and Matthew Colless$^{3}$\\
$^{1}$School of Physics, The University of Sydney, NSW 2006, Australia\\
$^{2}$Institute for Astronomy, Royal Observatory, Blackford Hill, Edinburgh EH9 3HJ, United Kingdom\\
$^{3}$Anglo-Australian Observatory, PO Box 296, Epping NSW 1710, Australia}
\begin{document}

\date{}

\pagerange{\pageref{firstpage}--\pageref{lastpage}} \pubyear{2005}

\maketitle

\label{firstpage}

\begin{abstract}
We investigate the topology of the completed 2dF Galaxy Redshift Survey, drawing two flux-limited samples of the local Universe from the 2dFGRS catalogue, which contains over 220,000 galaxies at a median redshift of $z=0.11$. The samples are cut at $z=0.2$ and corrected for selection effects. We then use the three-dimensional genus statistic to probe the connectedness of the galaxy distribution on scales ranging from 8 to 20 Mpc, and compare these measurements with the analytical result for a Gaussian random density field, a generic prediction of inflationary models. We demonstrate consistency with inflation on the range of scales considered. We then introduce a parameterisation of the analytic genus curve formula that is sensitive to asymmetries in genus number as a function of density and use it to demonstrate that such phenomena are ruled out with 95\% confidence between $8$ and $16$ Mpc.
\end{abstract}


\section{Introduction}\label{sec:intro}
Despite the success of the $\Lambda$CDM model in explaining the observations of cosmological surveys, the goal of understanding the formation and evolution of the largest structures in the Universe remains. Questions about the initial conditions of the Universe, from which structure evolves, can be answered by probing the largest scales with sufficient accuracy and density, and using methods of analysis that are adequately robust.

Topology has proved to be a valuable test of the properties of large-scale structure since the development of the genus statistic in the influential work of J. Richard Gott and his collaborators~\citep{GDM,Hamilton86,GWM,Gott89} and the more recent generalisation of this quantity~\citep[e.g.][]{Mecke94,Kerscher00}. The connectedness of the large-scale distribution of galaxies is used to probe the initial density field and the evolution of the peaks within the field, first as they grow linearly with the expansion of the Universe, and later non-linearly under gravitational clustering. The topological properties of this evolution have been investigated in analytical form~\citep{BBKS,Coles91}, where the non-linearity is frequently treated using a perturbation analysis~\citep[e.g.][]{Matsubara94,Matsubara96}, and tested against the observed galaxy distribution with many of the galaxy redshift surveys carried out to date~\citep{Vogeley94,Canavezes98}.

The genus statistic has often been used as a diagnostic to compare the primordial density field with that of Gaussian random phase conditions, a prediction of inflation~\citep{BST}. To date, no significant deviation from Gaussian initial conditions has been detected, though a common qualification is the wish for denser sampling over a larger volume.

The most recent galaxy redshift surveys, described in~\citet{Colless01} (the 2dF Galaxy Redshift Survey) and~\citet{Stoughton02} (the Sloan Digital Sky Survey), offer an opportunity to surmount this problem. Not only are they an order of magnitude larger in volume and in galaxy numbers, but the statistical methods used to fairly extract information from the raw data have increased in sophistication to the point where effectively all observed galaxies can be included without bias. The success of these surveys in analysing such statistical properties of the galaxy distribution as the power spectrum~\citep{Cole05}, void probability function~\citep{Croton04a} and the two-point and higher-order correlation functions~\citep{Hawkins03,Croton04b} gives great incentive for a topological analysis that will go significantly beyond previous work.

In this paper, we use the completed 2dF Galaxy Redshift Survey (2dFGRS) to examine the topology of the local universe. \citet{Hoyle2dF} have carried out a topological analysis of an early release 2dFGRS data set, using the two-dimensional genus statistic~\citep{Melott89} on a scale of 10 $h^{-1}$Mpc. We wish to extend this analysis to take advantage of the full public data release, using the three-dimensional genus statistic on scales between 8 and 20 $h^{-1}$Mpc. The choice to study a range of scales is an important one as it allows for the detection of a systematic variation in topology. We use a cosmology with $\Omega_m = 0.27$, $\Omega_\Lambda = 0.73$ and $h=0.72$. Hereafter we omit the presence of $h^{-1}$ when quoting distances in Mpc.

 Section~\ref{sec:survey} describes the data set and the preliminary analysis it is subject to in order to ensure statistical validity, before Section~\ref{sec:genus} reviews the topological genus and explains how it is calculated. Section~\ref{sec:results} then presents our results and the method of error analysis, discussing them within the context of the field.

\section[]{Redshift survey\\* samples}\label{sec:survey}
The Two-degree Field Galaxy Redshift Survey~\citep{Colless01} has measured the redshift of over 221 000 galaxies extending to a median redshift of $z=0.11$ and distributed mostly in two wedge-shaped slices. These form the independent, contiguous data sets that are analysed in this paper. The limiting magnitude of the survey in the $b_J$-band is 19.45 and the survey fields are tiled so as to provide a near-uniform sampling rate over the relevant regions of the sky. 

The raw data requires $k+e$ correction and the application of masks to account for variations in the limiting magnitudes and redshift completeness of the survey regions. These corrections are performed with the publically available software written by Peder Norberg and Shaun Cole, based on the calculation of the selection function in~\citet{Norberg02}. Information on the software, and the code itself, is provided on the survey website at \texttt{http://www2.aao.gov.au/2dFGRS/}.

\subsection{Selection function}
As with all redshift surveys, there is a need to account for redshift-dependent bias introduced by the fixed apparent magnitude limit. Previous studies of topology~\citep{Canavezes98,Hoyle2dF,Park05} have used samples constructed by selecting only objects above a fixed absolute magnitude cutoff (i.e. a volume-limited sample), effectively downweighting the over-represented population. We have chosen instead to use all the galaxies present in the NGP and SGP catalogues, and upweight those at high-redshift using the survey selection function.

\citet{Norberg02} have calculated the 2dFGRS selection function based on the $b_J$-band luminosity function, using the STY~\citep{Sandage79} and stepwise maximum-likelihood \citep[SWML;][]{Efstathiou88} estimators. The list of galaxy weights generated by the selection function is produced by the survey masking software. With a weighting value for each galaxy, we constructed purely apparent magnitude-limited samples from the survey data. 

\begin{figure}
  \centering
  \begin{minipage}[c]{.48\textwidth}
    \centering
    \includegraphics[width=78mm]{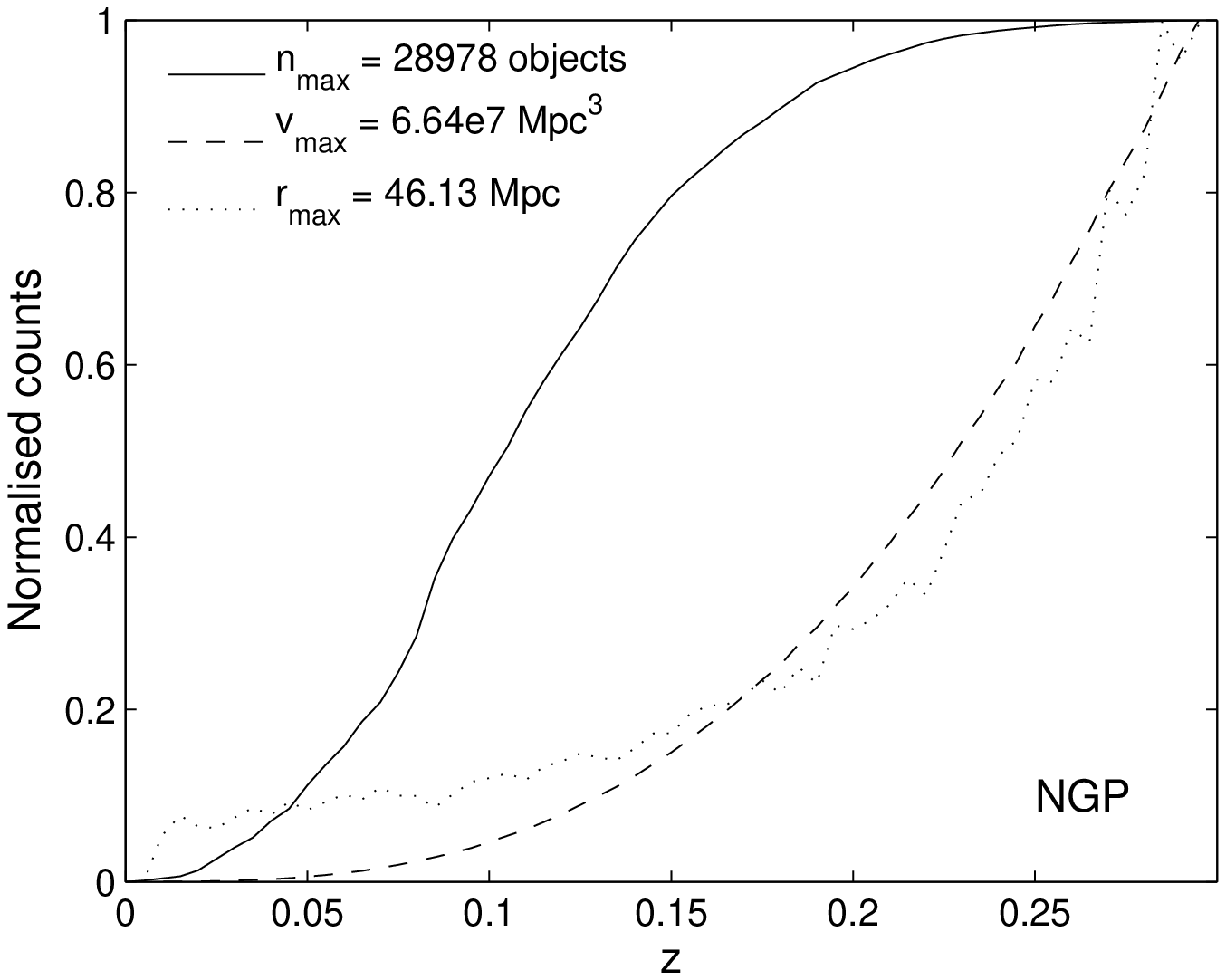}
  \end{minipage}
  \hspace{0.2cm}
  \begin{minipage}[c]{.48\textwidth}
    \centering
    \includegraphics[width=78mm]{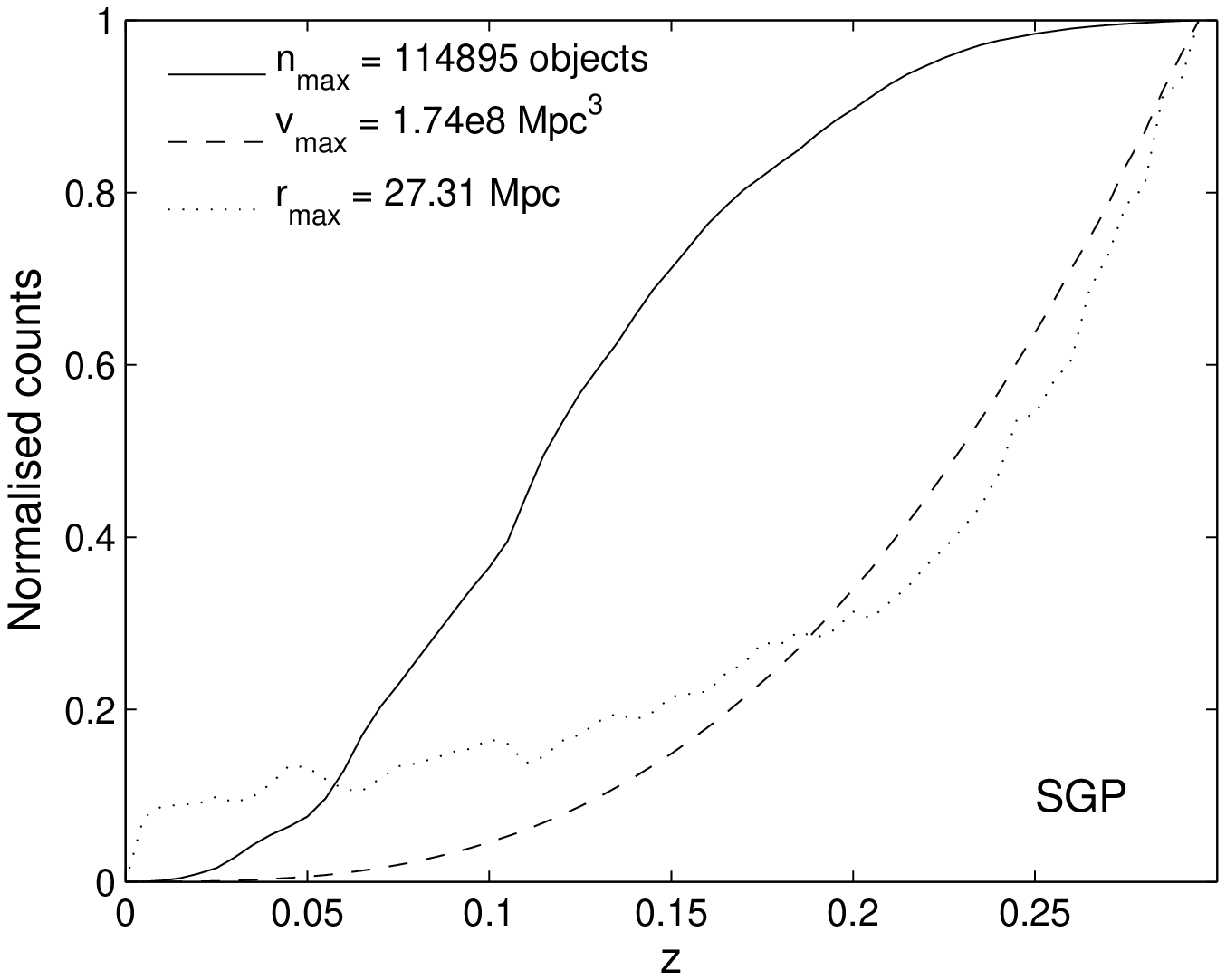}
  \end{minipage}
  \caption{Object counts for both NGP and SGP samples, along with the volume of the survey and an estimate of the mean inter-galaxy separation. The value of $r$ is calculated incrementally, as there will be systematically less points (though they have higher weightings) at larger redshift. It is this maximum incremental value of the smoothing length that needs to be used in order the represent the data fairly.} \label{fig:sf}
\end{figure}

Figure~\ref{fig:sf} shows the sample size, sample volume and estimated mean inter-galaxy separation as a function of redshift for both the NGP and SGP slices. This estimate of the sampling density is found by computing the inverse number density of objects in the survey region --- i.e.~the volume per object --- and finding the radius of the sphere about each galaxy that would contain this volume.
\begin{equation} \alpha = \frac{V}{N} =\frac{4}{3}\pi r^3 \Rightarrow r = \left(\frac{3\alpha}{4\pi}\right)^{1/3},\label{eq:migs} \end{equation}
where $N$ is the total number of objects in the sample, and $V$ is the number of survey cells multiplied by the volume of each cell ($4^3$ Mpc). Significantly, there is a break in the value of $r$ near $z\sim0.2$, corresponding to a drop in the spatial completeness of the \emph{survey}, that represents a natural choice of maximum sample size.

\subsection{Smoothing the Samples}\label{sec:smooth}
Once the two correctly weighted point distributions have been produced, it remains to turn them into a continuous field amenable to a topological measurement. A natural choice of smoothing kernel is a Gaussian, as it minimises the effect of high-frequency `ringing' that occurs when sharp cutoffs are present. Moreover, the form of the genus curve of a field smoothed through a Gaussian window has been calculated by~\citet{Hamilton86}.

\citet{Martinez05} have argued that Gaussian smoothing is restrictive in the sense that there is fixed range of smoothing lengths, outside of which properties of the smoothing dominate over the signal. They advocate instead a multiscale smoothing process, in particular wavelet analysis. We suggest that the selection feature of a fixed-scale smoothing process is in fact a desirable characteristic, as the comparison of topology on different scales provides information that is important for cosmology.

The transition between the regimes of linear and non-linear structure is smooth and very broad in scale. A characteristic length occurs where the fractional RMS amplitude of density perturbations exceeds unity. This is represented by the cosmological parameter $\sigma_n$, the fractional RMS amplitude between spheres of radius $n$ Mpc, with $n=8$ frequently used as $\sigma_8$ is not too far from unity. Indeed, the present determination of $\sigma_8$ from several independent sources~\citep[e.g.][]{Tytler04} has converged near 0.9. We have used the value $\sigma_8 =0.89$, in line with the clustering analysis of this data by 2dFGRS team members. This suggests that the characteristic smoothing scale is $\sim10$ Mpc.

\begin{figure}
  \centering
  \begin{minipage}[c]{.48\textwidth}
    \centering
    \includegraphics[width=78mm]{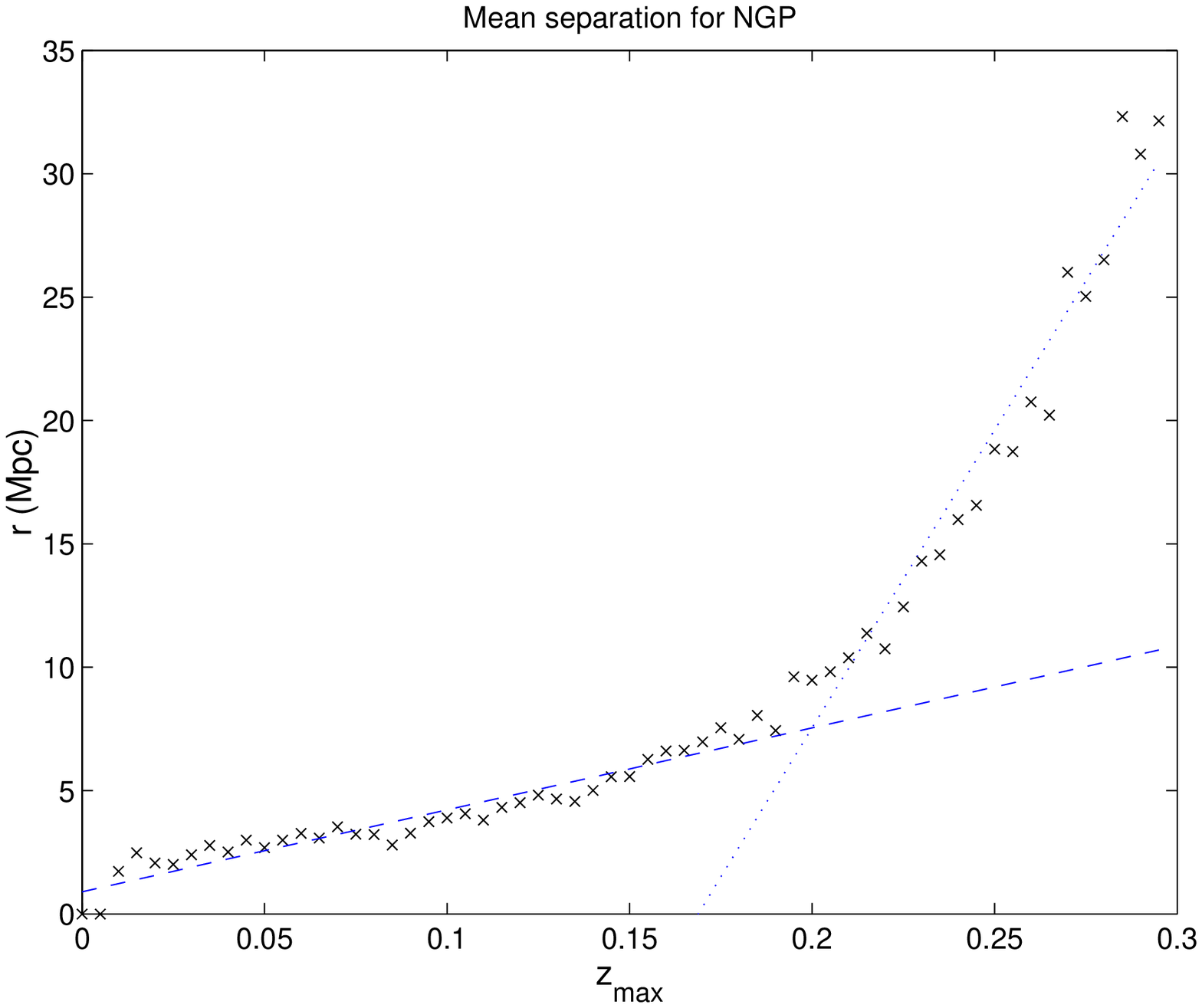}
  \end{minipage}
  \hspace{0.2cm}
  \begin{minipage}[c]{.48\textwidth}
    \centering
    \includegraphics[width=78mm]{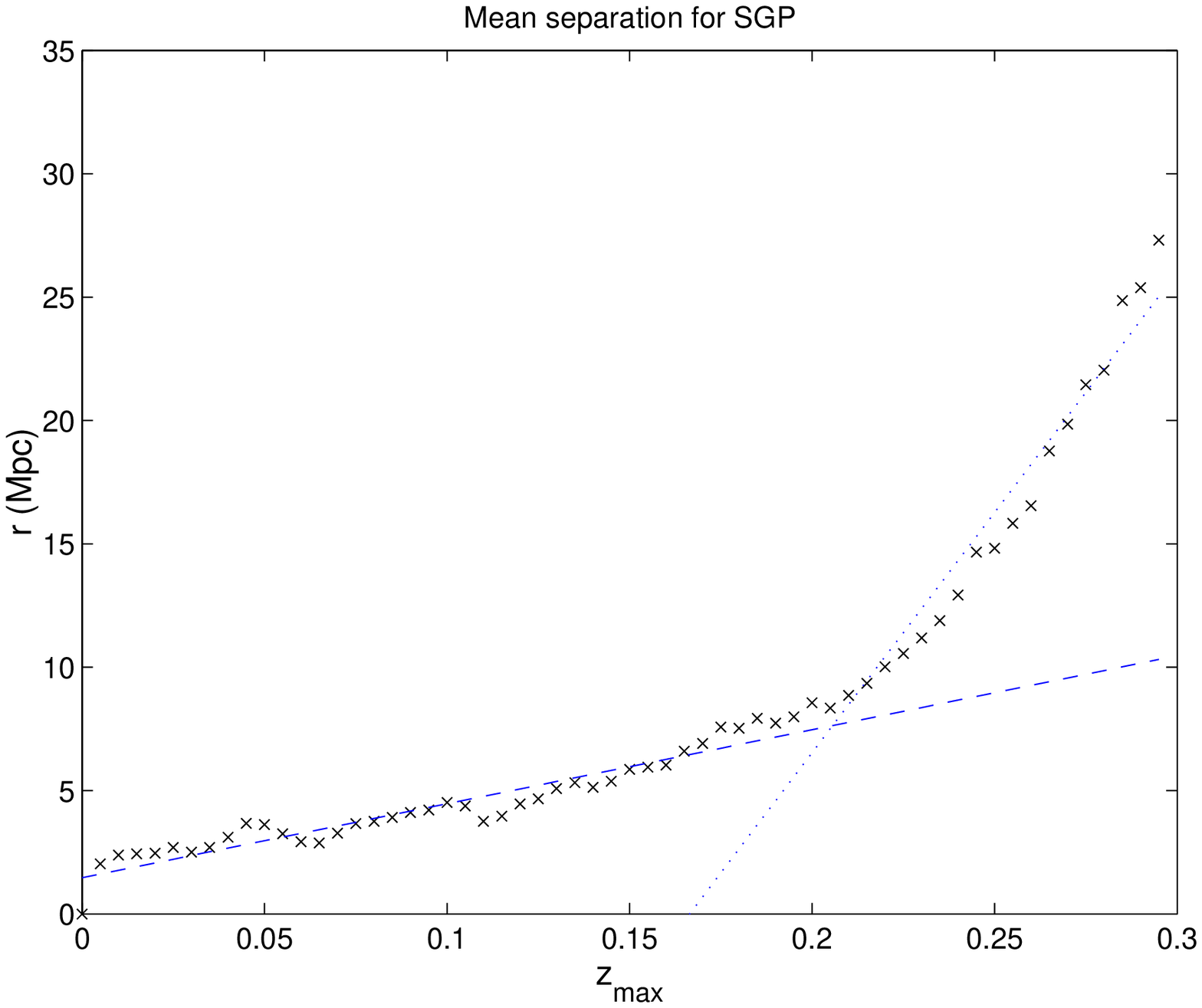}
  \end{minipage}
  \caption{Mean inter-galaxy separation $r$ as a function of maximum sample redshift. The break near $z\sim0.2$ for both slices is the prominent feature. As more galaxies were targeted in the SGP slice, the values of $r$ are somewhat lower than for the NGP.} \label{fig:slopes}
\end{figure}

An important consideration is that the separation of points in the field be well below the smoothing length. The value of $r$, defined in Equation~\ref{eq:migs}, for objects at a given redshift, provides an upper bound on this length. These values are shown in Figure~\ref{fig:slopes}. The break is illustrated by fitting two straight lines to different regions of the plot. 
The value of $r$ for a choice of $z_{max}$ is a good guide to the smoothing length that will result in a true \emph{field}, i.e.~not a distribution that, while continuous, is really still a series of island peaks in an otherwise shallow sea. We decided therefore that as it was worthwhile exploring structures on a range of scales, selecting 8 Mpc as a lower bound for the smoothing length, extending up to only 20 Mpc, in steps of 2 Mpc. To examine the topology in this range requires that the choice of $z_{max}$ be restricted so as to allow smoothing down to $\sim8$-$10$ Mpc, i.e. $z_{max}\sim0.2$.

The non-periodic geometries of the 2dF NGP and SGP regions permit edge effects in the form of a power loss near the survey boundaries, resulting in a systematic underdensities that have a pronounced effect on topological measurements. To undo this, the method described by~\citet{Melott93} was followed, whereby the edge effect is emulated in a separate field of constant density, and then ratioed out. To do this, a field with the same geometry as the survey but of constant positive density (set to zero elsewhere) is subject to the same smoothing process as the data. The smoothed data field is then ratioed, on a cellwise basis, with the smoothed constant field. Conceptually, the edge effect has been factored out, at the cost of adding some noise to the topological information near the survey boundaries. 

The volume affected by this process can be estimated by comparing the power inside the survey region before and after smoothing. This is a conservative estimate in the sense that even cells with small fractional changes in density, well away from the edges of the array, will count toward the power transmitted outside the survey region - it is akin to measuring the `equivalent volume' of the smoothing process. This transmission increases linearly with smoothing scale; for the range of scales we consider, the volume affected for the NGP region is 12.7\% (smoothing over 8 Mpc) and 26.8\% (over 20 Mpc), and for the larger SGP slice is 10.0\% (8 Mpc) and 20.7\% (20 Mpc).

\section{Topological Analysis}\label{sec:genus}
At the heart of the topological analysis is the computation of the genus-threshold density relation, described in ~\citet{Gott89}. The method calculates the genus number of a sequence of two-dimensional surfaces embedded in the three-dimensional sample volume. The surfaces are defined as the boundaries between the regions above and below a fixed critical density, parameterised by the quantity $\nu$. Each value of $\nu$ excises a monotonically increasing volume of the sample. The fractional volume enclosed by the surface as a function of $\nu$ is
\begin{equation} v_f(\nu) = \frac{1}{\sqrt{2\pi}}\int_\nu^\infty e^{-t^2/2}dt = \frac{1}{2}\textrm{Erf}_c\left(\frac{\nu}{\sqrt{2}}\right), \label{eq:nu}\end{equation}
where Erf$_c$ is the conjugate error function. This number is used to find the critical density of the distribution, $\rho_c$. In the CONTOUR3D code~\citep{Weinberg88}, a Newton-Raphson iteration process is used to determine this value. We have instead chosen to compute the critical density by sorting the three-dimensional density distibution into a one-dimensional array ordered by increasing density, so that, as each element of the array represents a standard unit of volume, moving $n$\% of the way along the array will give the critical density corresponding to enclosing $(100-n)$\% of the volume. That is, $\rho_c$ is the $m$th element of the sorted arrays, for
\begin{equation} m = (1-v_f)N, \end{equation}
where $N$ is the number of elements in the arrays ($N$ being the same for both). A similar method has been employed previously by~\citet{Canavezes98}.

The genus number of each two-dimensional surface is then measured using the standard CONTOUR3D algorithm, with, in particular, the same angle deficit matrix for each of the staggered configurations, though we have written our own code in Fortran 90 to calculate this. Those cells above the critical density are labelled with the value $1$; those below are labelled $0$; those outside the survey region, as defined by the range in right ascenion and declination in~\citet{Colless01} are labelled $-256$. This produced a polygonal surface on which the Gaussian curvature is compressed into the vertices of the polygons. As is conventional, we define $g_s$ in terms of the integrated Gaussian curvature $K$,
\begin{equation} g_s = g - 1 = -\frac{1}{4\pi}\int KdA = -\frac{1}{4\pi}\sum D_i,\end{equation}
for the angle deficit defined by $D_i = 2\pi-\sum_i V_i$, where each of the $V_i$ are the angles around a vertex. 

~\citet{BBKS} have derived the genus number of isosurfaces through a Gaussian random field, parameterised by the quantity $\nu$ described in Equation~\ref{eq:nu},
\begin{equation} g(\nu;N) = N(1-\nu^2)e^{-\nu^2/2}, \label{eq:genus}\end{equation}
where the normalisation factor $N$ depends on the exact choice of the power spectrum of the field. For a featureless power law of spectral index $n$, smoothed by a Gaussian of width $\lambda$, the normaliation factor is~\citep{MartinezSaar}

\begin{equation}
N = \frac{1}{4\pi^2}\left(\frac{\langle k^2\rangle}{3}\right)^{3/2} = \frac{1}{4\pi^2\lambda^3}\left(\frac{3+n}{3}\right)^{3/2}
\end{equation}

In practice, the genus curve of the cosmological density field has agreed well with this model to first-order. However, the curve is sufficiently variable that a pure Gaussian random field may be inadequate to give a wholly satisfatory description. Indeed it is entirely sensible to consider an alternative model built upon the foundation of a Gaussian random field, but differing in parametric properties that are mutually independent and open to physical interpretation. Motivated by consideration of the moments of the normal distribution, define a `standardisation' of the genus curve
\begin{equation} g_s(\nu^\prime;\mathbf{P}_3) = N\left(1 - \nu^{\prime2}\right)\exp\left(-\nu^{\prime2}/2\right), \end{equation}
where
\begin{equation} \nu^\prime = \frac{\nu - \mu}{\sigma}, \end{equation}
and $\mathbf{P}_3 = (N,\mu,\sigma)$ is a three-parameter vector describing the transformation. Modifications from $\mathbf{P}_3 = (N,0,1)$ will recentre ($\mu$) and stretch ($\sigma$) the genus curve. As a further probe of this, we define a fourth parameter $\alpha$ to describe the imbalance between the two minima of the genus curve, increasing (for positive $\alpha$) the relative abundance of connected high-density clumps (i.e. a `meatball' topology, as \emph{per}~\citet{So78}),
\begin{equation} g_s(\nu^\prime;\mathbf{P}_4) = N\left(1 - \nu^{\prime2}\right)\exp\left\{-(\nu^{\prime2} - \alpha\nu^\prime)/2\right\}. \end{equation}
Each model is now described by the analogous four-parameter vector $\mathbf{P}_4 = (N,\mu,\sigma,\alpha)$, and the class of Gaussian random fields is $\mathbf{P}_4 = (N,0,1,0)$. 

The determination of the components of $\mathbf{P}_4$ proceeds by minimising the reduced $\chi^2$ function
\begin{equation} \chi^2(N,\mu,\sigma,\alpha) = \frac{1}{n-k}\sum_{i=1}^n\left(\frac{y_i - g(\nu_i,N,\mu,\sigma,\alpha)}{\sigma_i}\right)^2, \end{equation}
where $y_i$ and $\sigma_i$ are the data points and their error bars respectively, $g(\nu_i;N,\mu,\sigma,\alpha)$ is the four-parameter model\footnote{The minimisation is performed using the Numerical Recipes Fortran 77 \texttt{amoeba} routine~\citep{NR}.}. Here $n-k$ represents the number of degrees-of-freedom (the number of independent points less the number of parameters in the model).

For the range of $\nu$ over which this curve is fit, there are 51 data points --- as these define surfaces through the same density field, they cannot be fully independent, so the value of $n$ in the reduced $\chi^2$ is smaller than the number of data points used to fit the model. Moreover, increasing this number by measuring over a finer grid of $\nu$ will not increase the information contained in the data. It is possible to estimate the true number of degrees-of-freedom by measuring the unreduced $\chi^2$ values of the analytic fits (Eq.~\eqref{eq:genus}) to the genus measurements from realisations of a Gaussian random field. The mean value of this sample of the $\chi^2$ distribution is an unbiased estimator of the number of degrees-of-freedom. Using 10 realisations of a $128^3$ Gaussian random field, smoothed with a Gaussian kernel of width $8$ Mpc, and a genus curve with 50 points, the number of degrees-of-freedom is $45\pm7.0$ (the 1$\sigma$ range is generated by bootstrap resampling). Each of these fits leaves the amplitude of the genus curve as a free parameter --- for the four-parameter model, we subtract one for each of the remaining parameters. This defines the effective number of degrees-of-freedom to be used in calculating the reduced-$\chi^2$.

\section{Results}\label{sec:results}
Inflation predicts that the universe will be a Gaussian random field on scales unaffected by gravitationally-induced non-linear structure formation~\citep{BST}. If the universe is of infinite extent and of finite age, it is a matter of looking to large enough scales to see this reflected in the galaxy distribution. 

\begin{figure*}
\centering
\includegraphics[width=185mm]{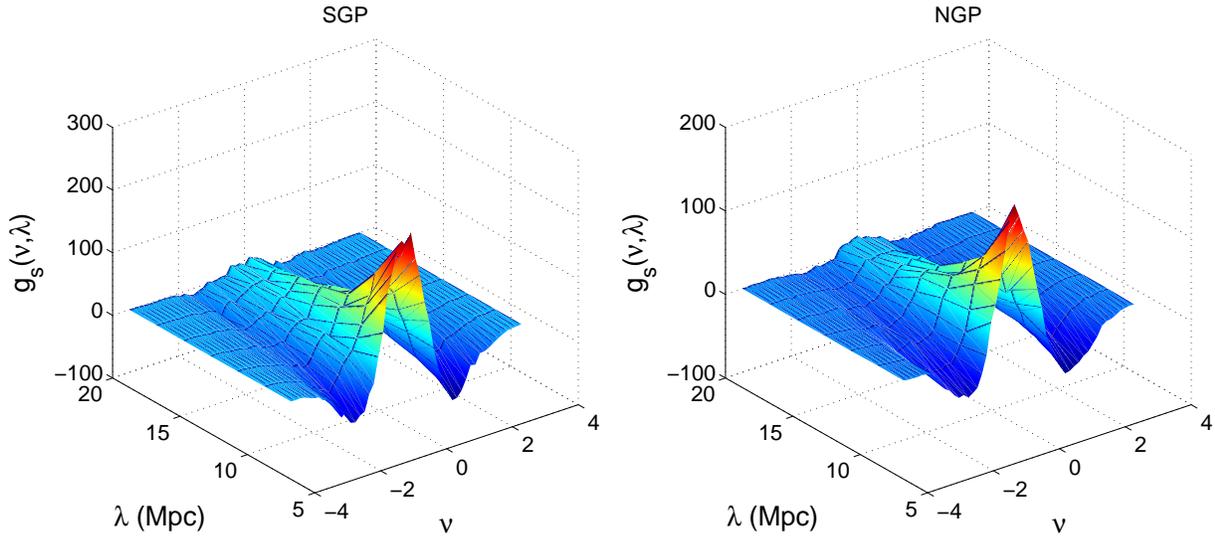}
\caption{Genus surfaces for the two slices in the 2dFGRS. Both exhibit an apparent asymmetry at small values of $\lambda$; the degree to which this is the case at larger scales is unclear as a result of the decline in amplitude with scale length --- an effect that is part of the smoothing, rather than representative of a change in the power spectrum with scale.}\label{fig:2df_surf}
\end{figure*}

\begin{figure*}
\centering
\includegraphics[width=185mm]{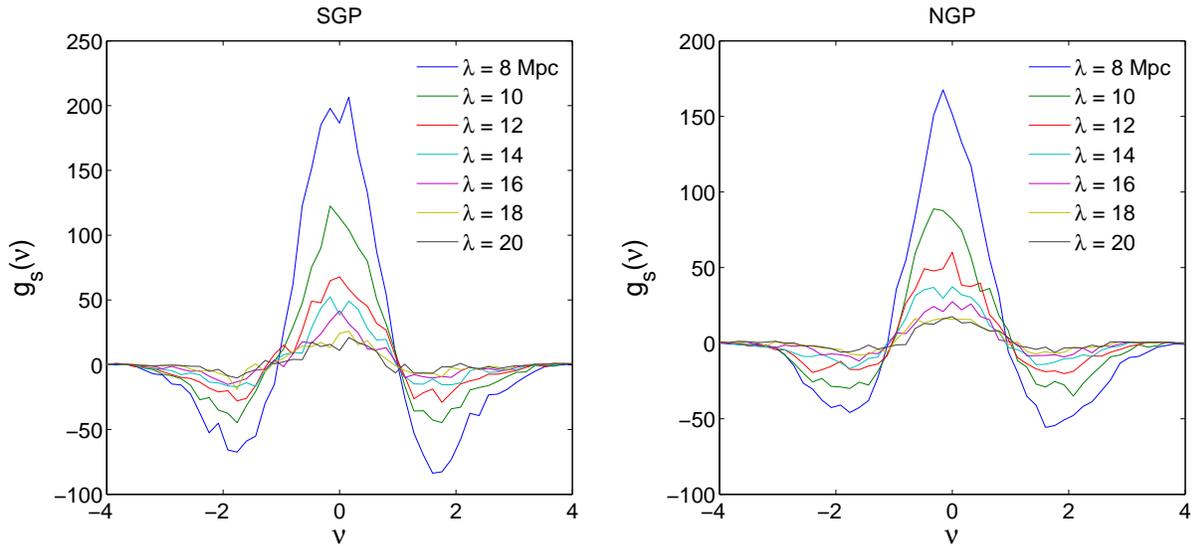}
\caption{Slices through the genus surfaces for the 2dFGRS. Both show the characteristic asymmetry of a meatball topology on small scales, but the trend with scale toward or away from a Gaussian random field is not clear by comparing the curves in this relative fashion.}\label{fig:2df_slice}
\end{figure*}

The genus surfaces for the SGP and NGP slices, constructed from flux-limited samples, are shown in Figure~\ref{fig:2df_surf}.
These two surfaces, which represent large ($\sim10^7$ Mpc$^3$), independent regions of the universe, are in good agreement. They appear to agree with Gaussian random field topology, exhibiting the expected symmetry properties and scale-dependence of amplitude. A departure from symmetry is apparent on small scales, where the amplitude of the genus curves is largest. The asymmetry in the regions of negative genus (`troughs') is indicative of a greater number of isolated high-density regions than isolated low-density regions. On larger scales, the asymmetry is not apparent, though in the surface representation it is unclear whether this is the result of the characteristic reduction in amplitude with scale. Figure~\ref{fig:2df_slice} is a plot of slices through this surface at values of $\lambda$ between 8 and 20 Mpc, in steps of 2 Mpc.

\begin{figure*}
\centering
\begin{minipage}[c]{0.48\textwidth}
\centering
\includegraphics[width=78mm]{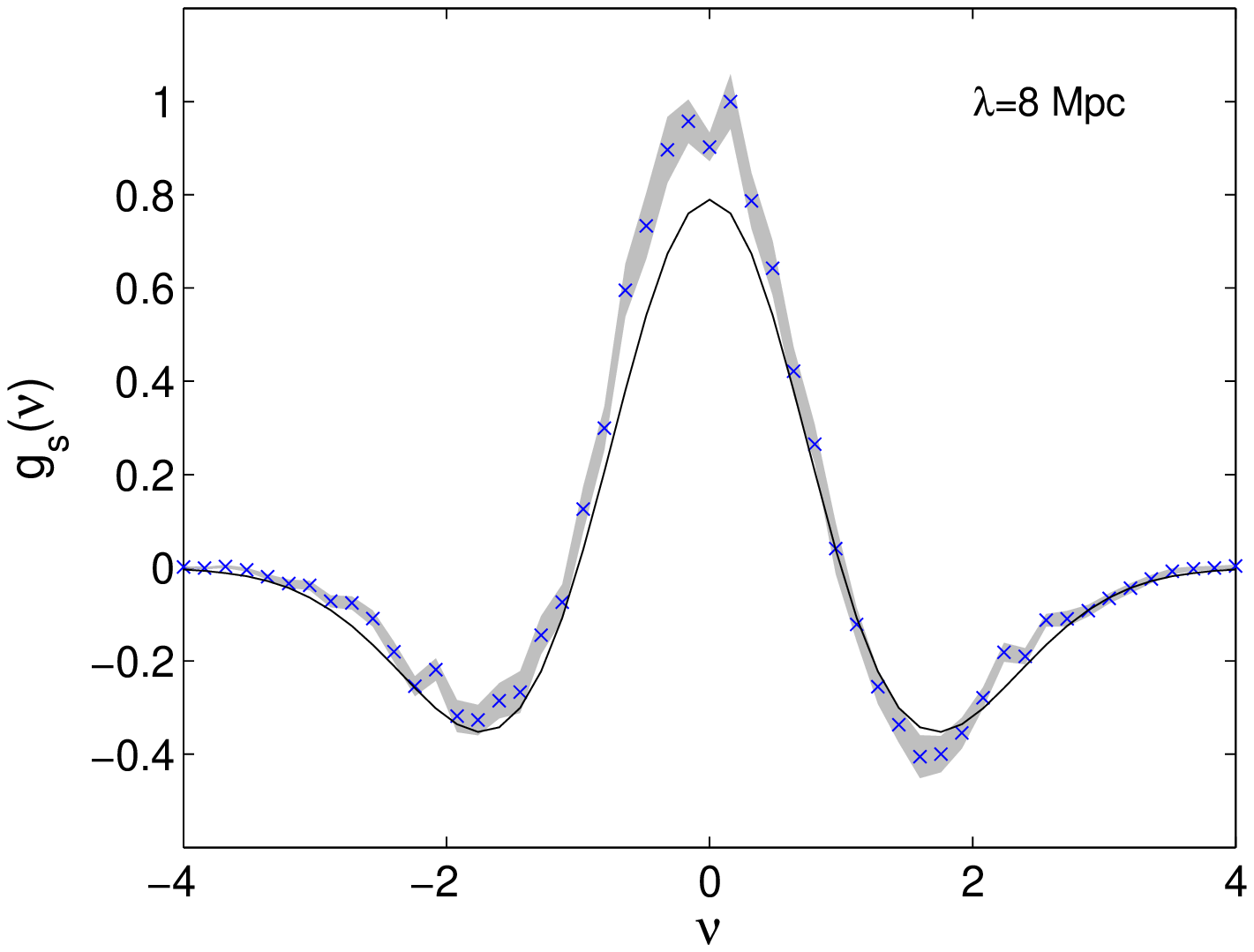}
\end{minipage}
\hspace{0.2cm}
\begin{minipage}[c]{0.48\textwidth}
\centering
\includegraphics[width=78mm]{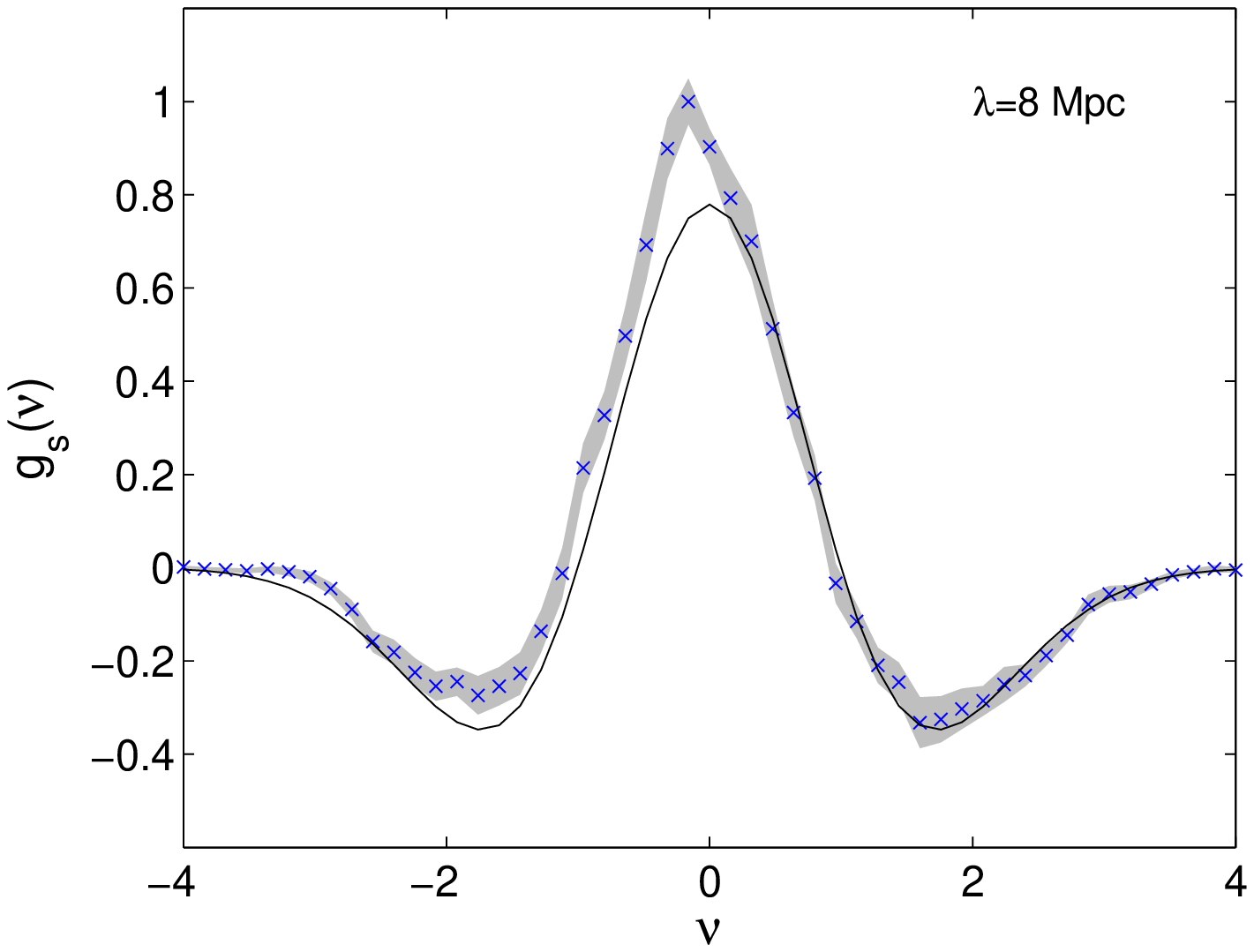}
\end{minipage}
\vspace{0.5cm}
\begin{minipage}[c]{0.48\textwidth}
\centering
\includegraphics[width=78mm]{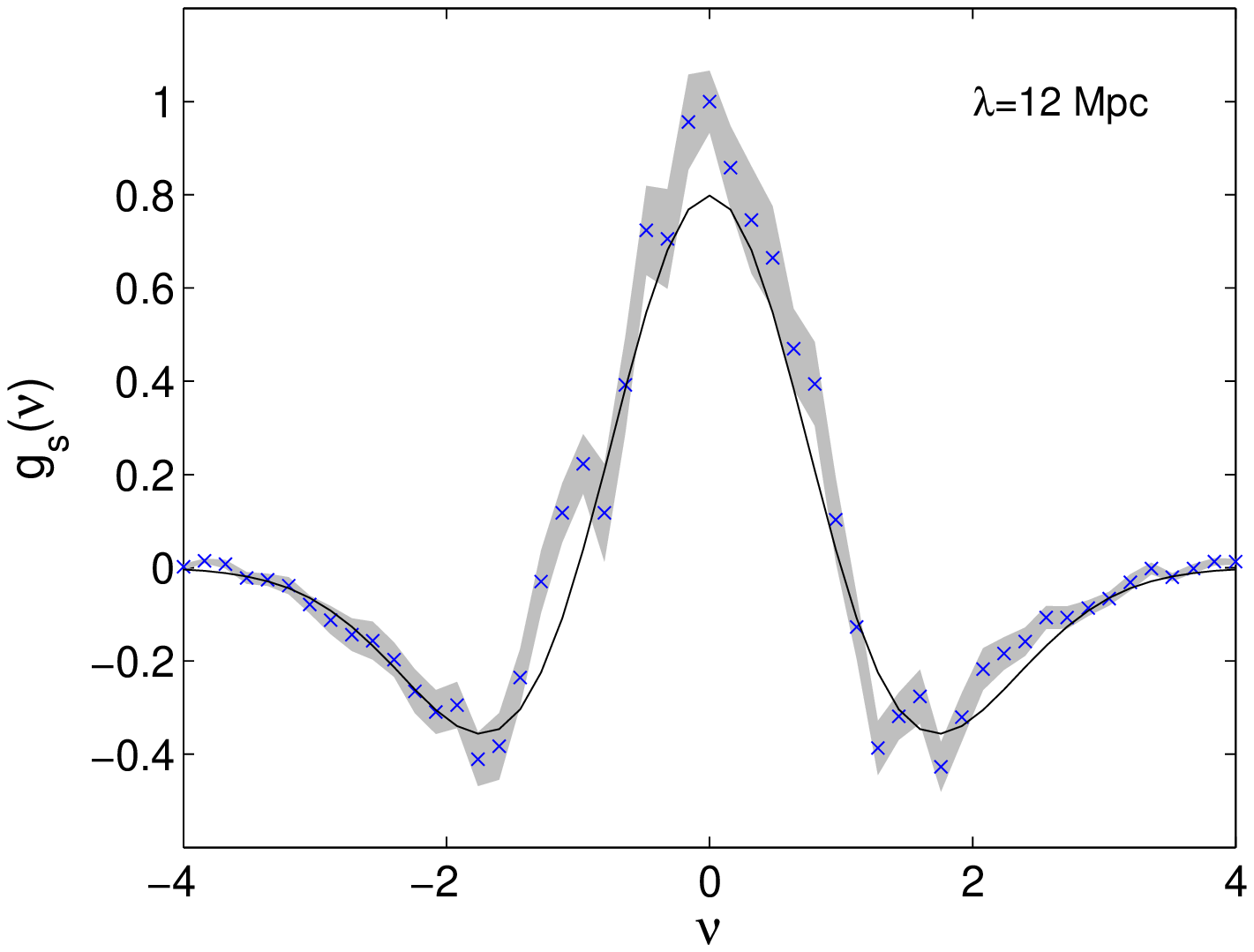}
\end{minipage}
\hspace{0.2cm}
\begin{minipage}[c]{0.48\textwidth}
\centering
\includegraphics[width=78mm]{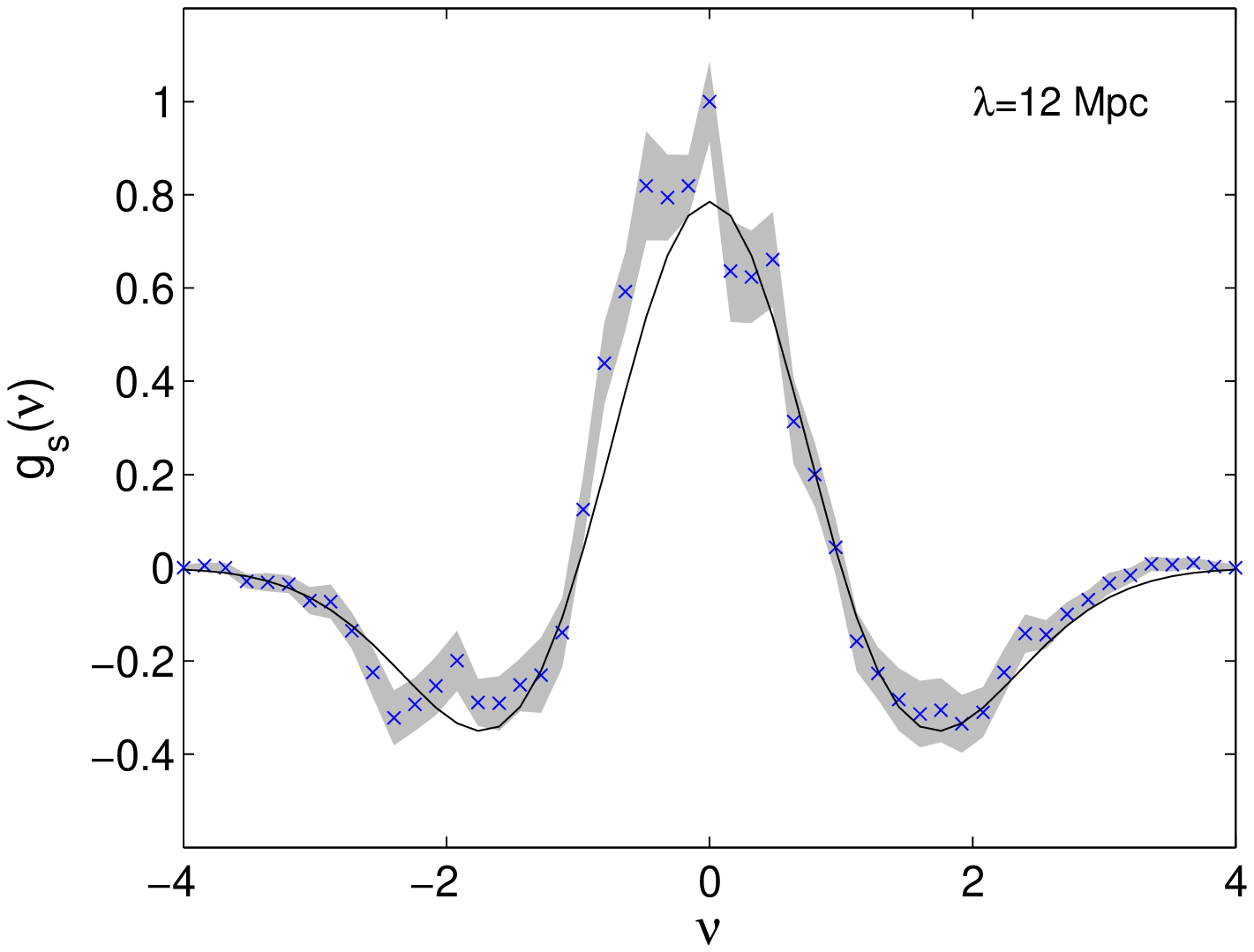}
\end{minipage}
\vspace{0.5cm}
\begin{minipage}[c]{0.48\textwidth}
\centering
\includegraphics[width=78mm]{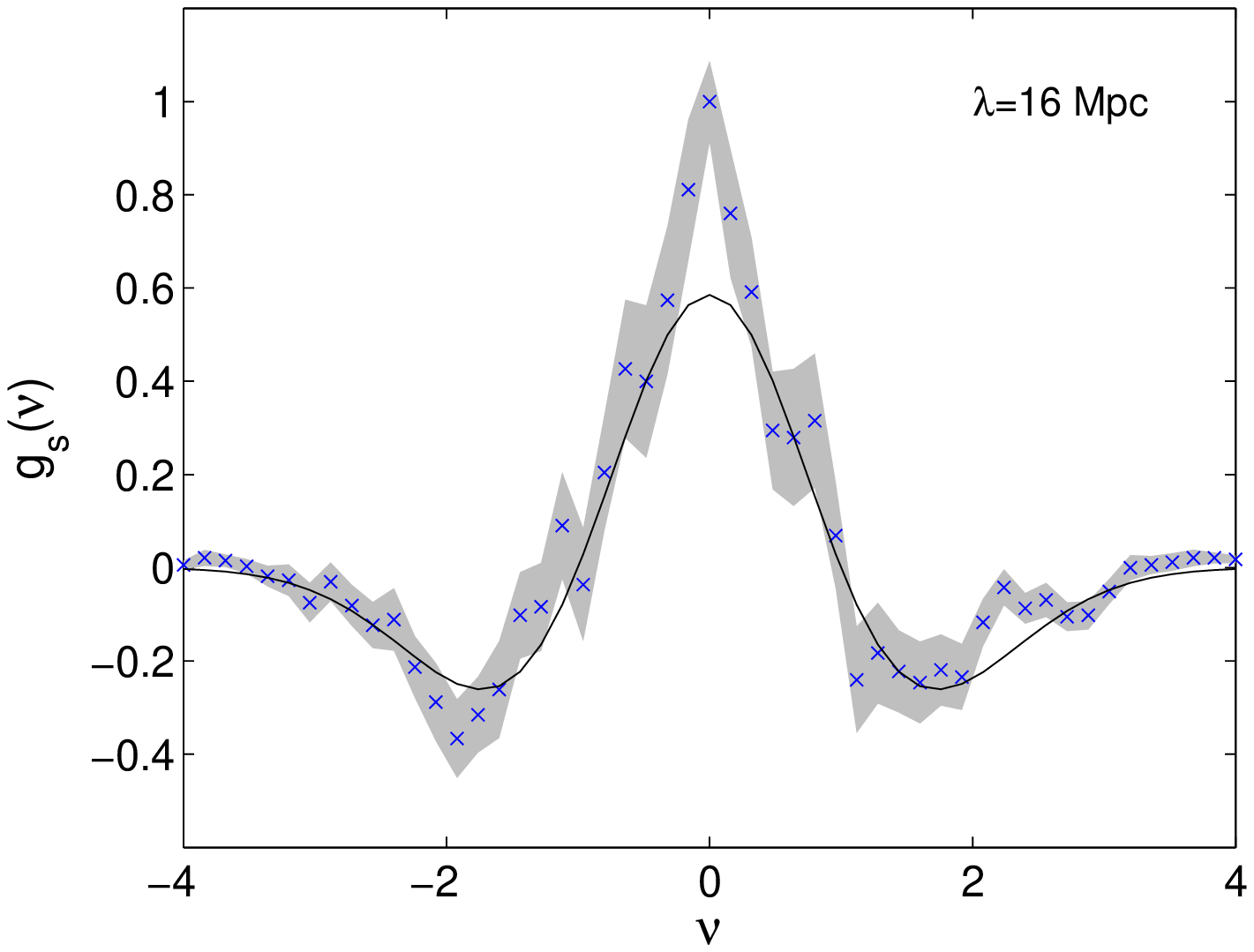}
\end{minipage}
\hspace{0.2cm}
\begin{minipage}[c]{0.48\textwidth}
\centering
\includegraphics[width=78mm]{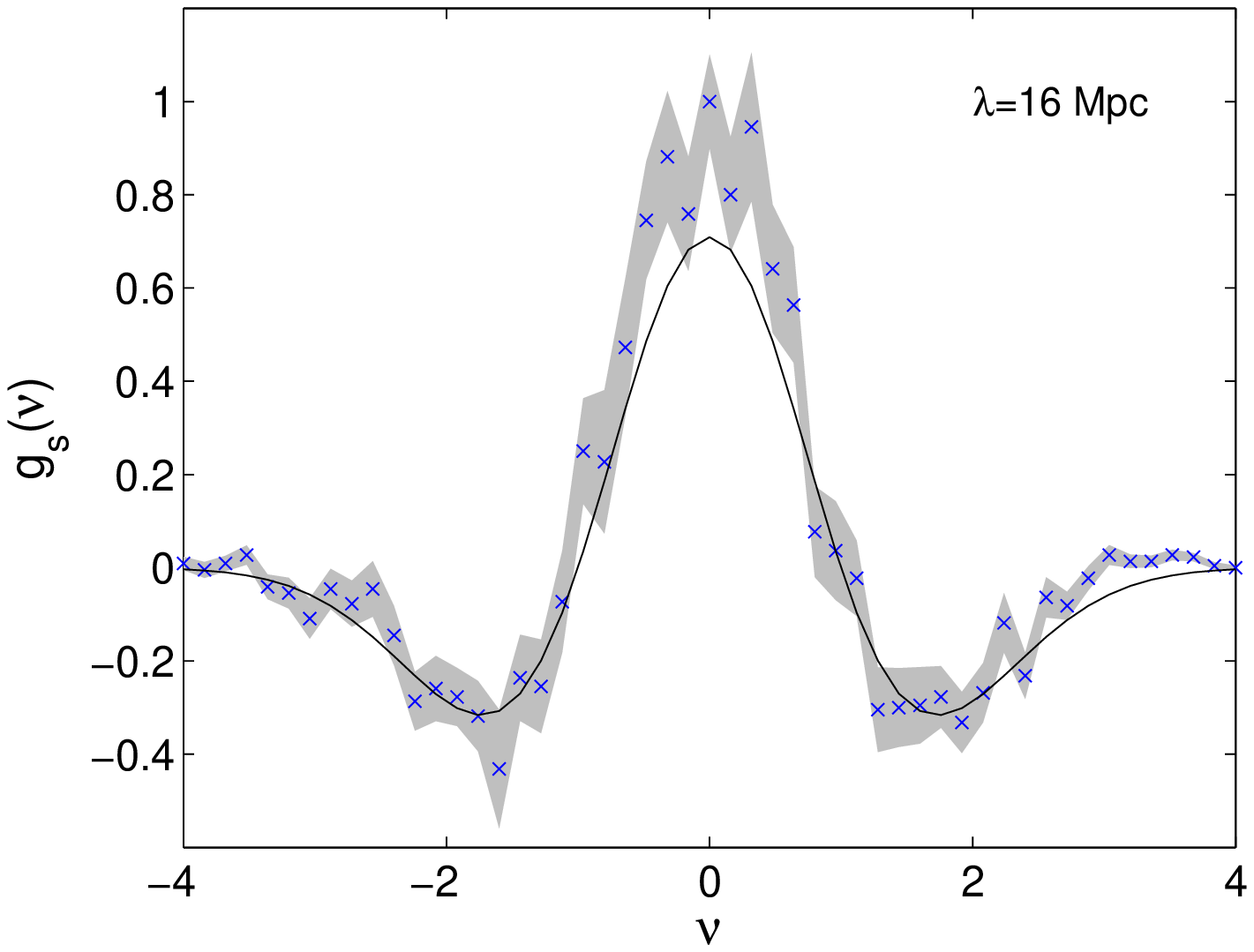}
\end{minipage}
\caption{Genus curves for the SGP (left) and NGP slices at $\lambda = 8$ (top), $12$ and $16$ Mpc. The data points are the crosses with $1\sigma$ error bars calculated using a bootstrap resampling method. The solid line is the best fit model genus curve for a pure Gaussian random field, with the amplitude as a free parameter as described in the body text. The fit is a minimum in reduced $\chi^2$ over the region $-3.5\le\nu\le3.5$.}\label{fig:2df_genus}
\end{figure*}

\begin{figure*}
\centering
\begin{minipage}[c]{0.48\textwidth}
\centering
\includegraphics[width=78mm]{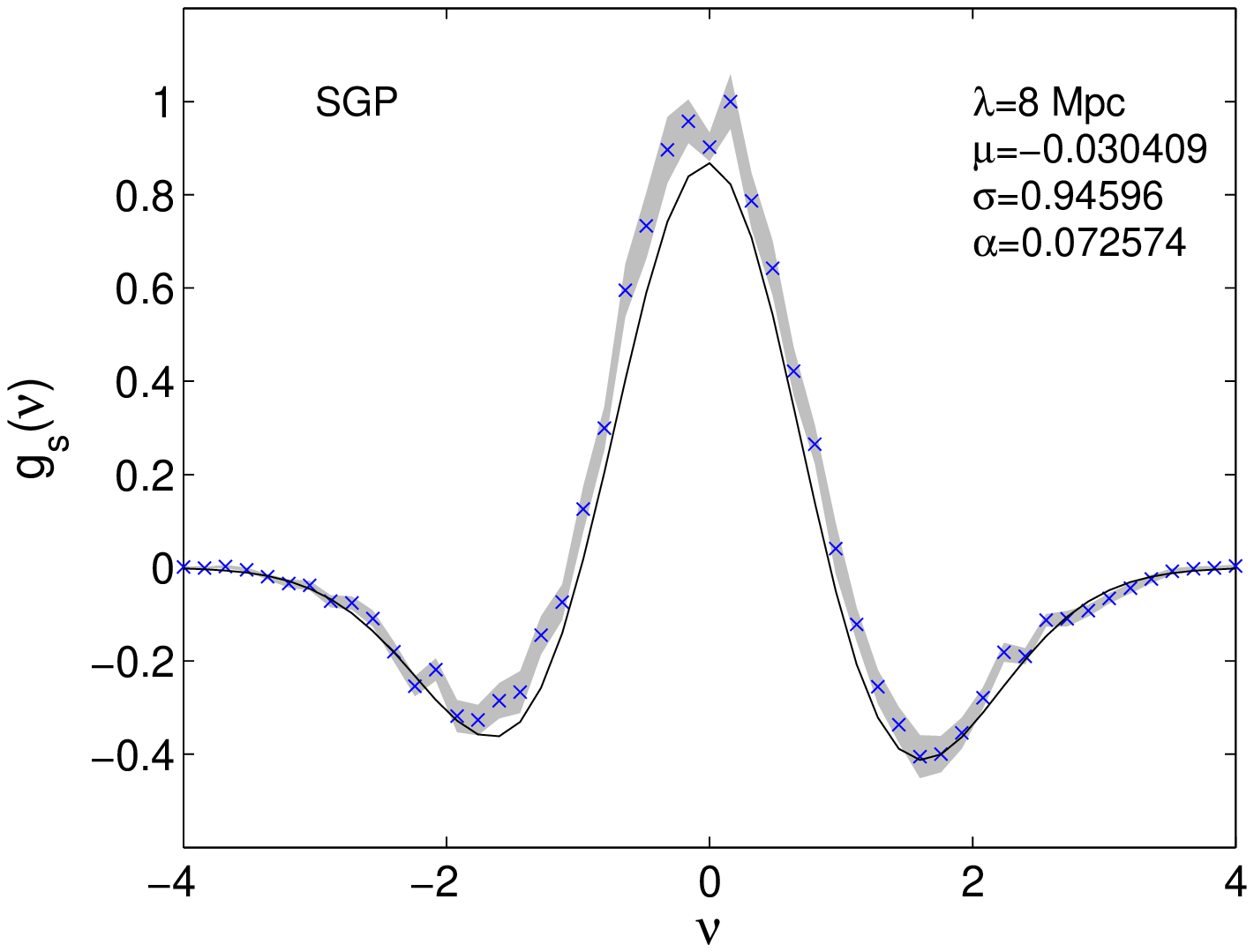}
\end{minipage}
\hspace{0.2cm}
\begin{minipage}[c]{0.48\textwidth}
\centering
\includegraphics[width=78mm]{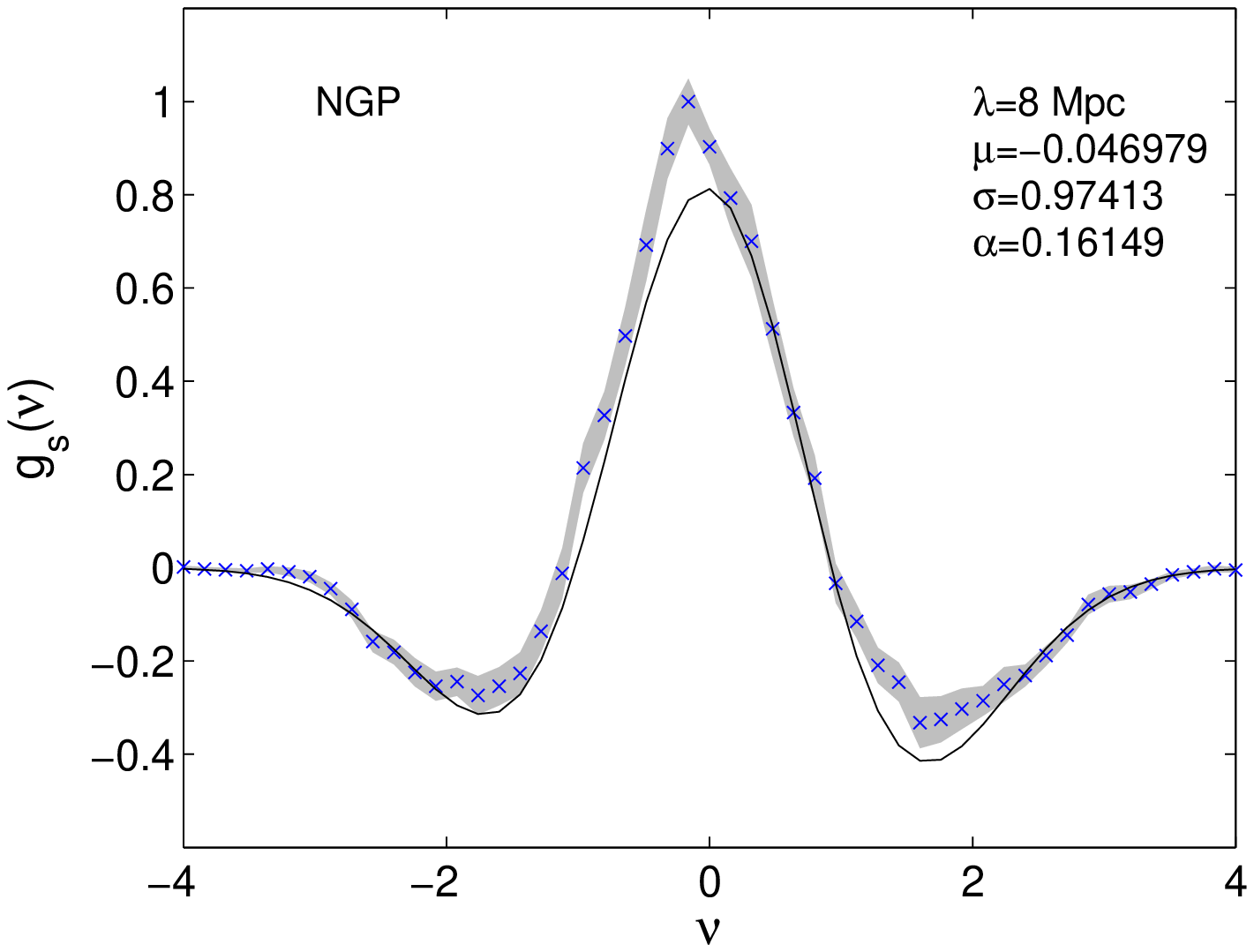}
\end{minipage}
\vspace{0.5cm}
\begin{minipage}[c]{0.48\textwidth}
\centering
\includegraphics[width=78mm]{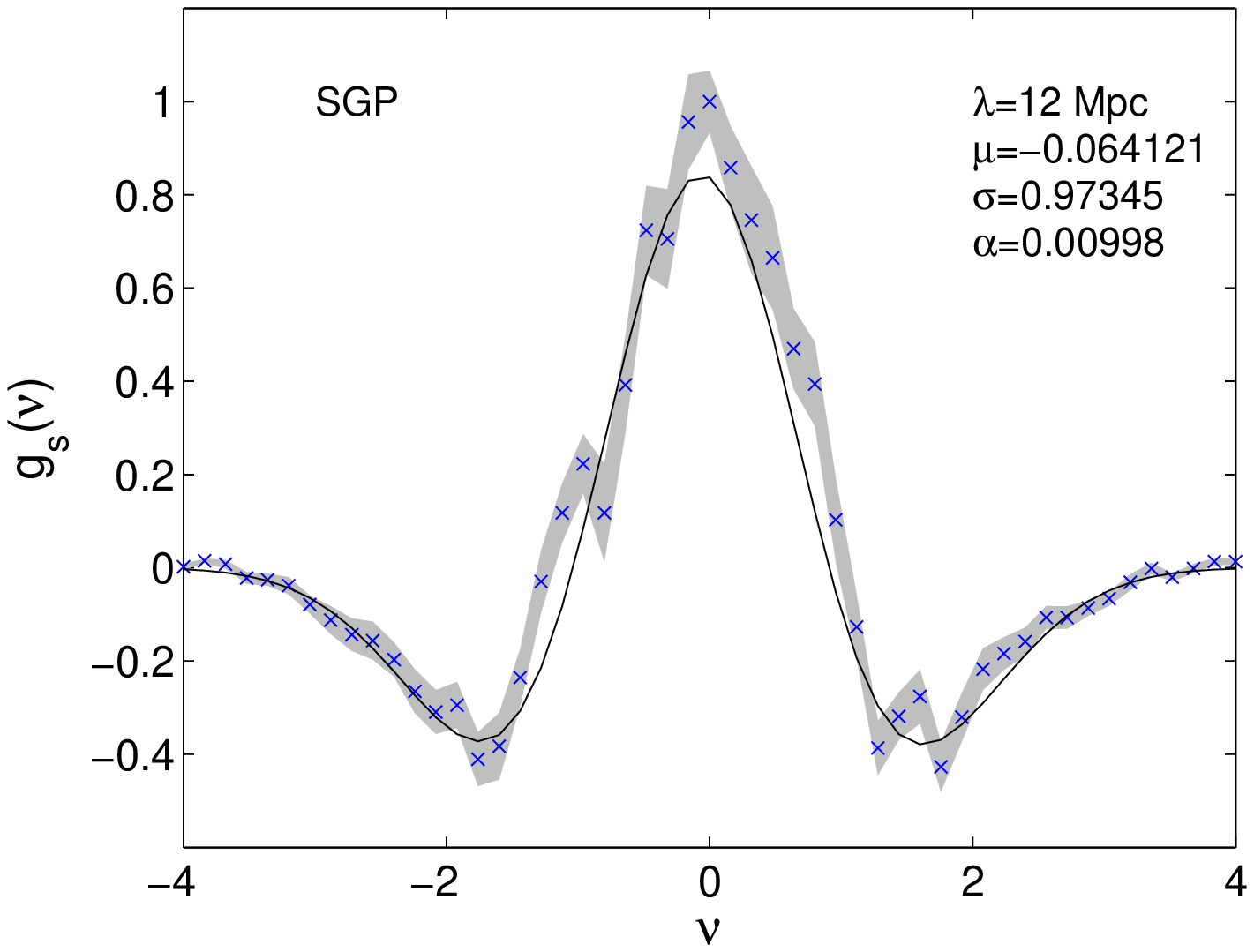}
\end{minipage}
\hspace{0.2cm}
\begin{minipage}[c]{0.48\textwidth}
\centering
\includegraphics[width=78mm]{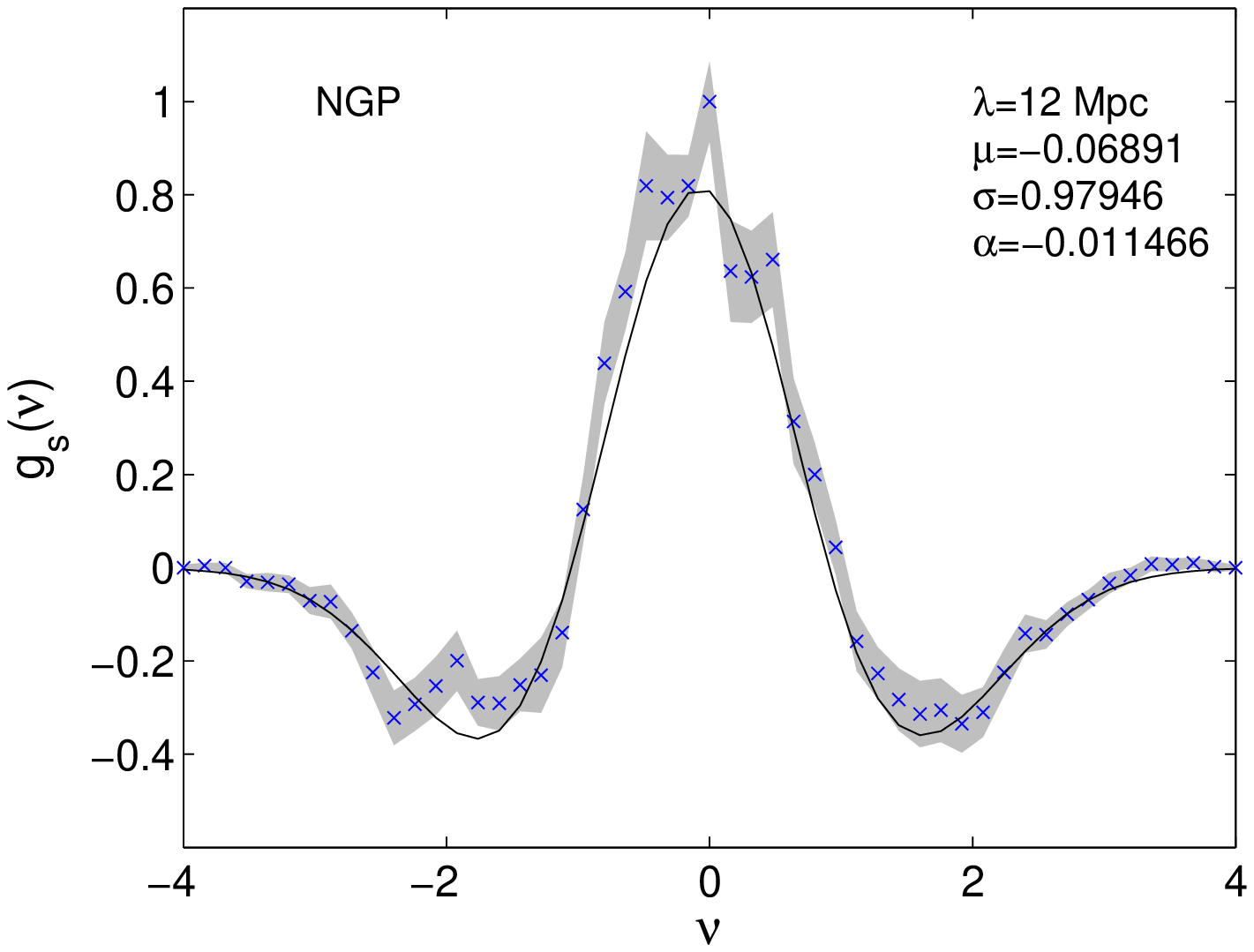}
\end{minipage}
\vspace{0.5cm}
\begin{minipage}[c]{0.48\textwidth}
\centering
\includegraphics[width=78mm]{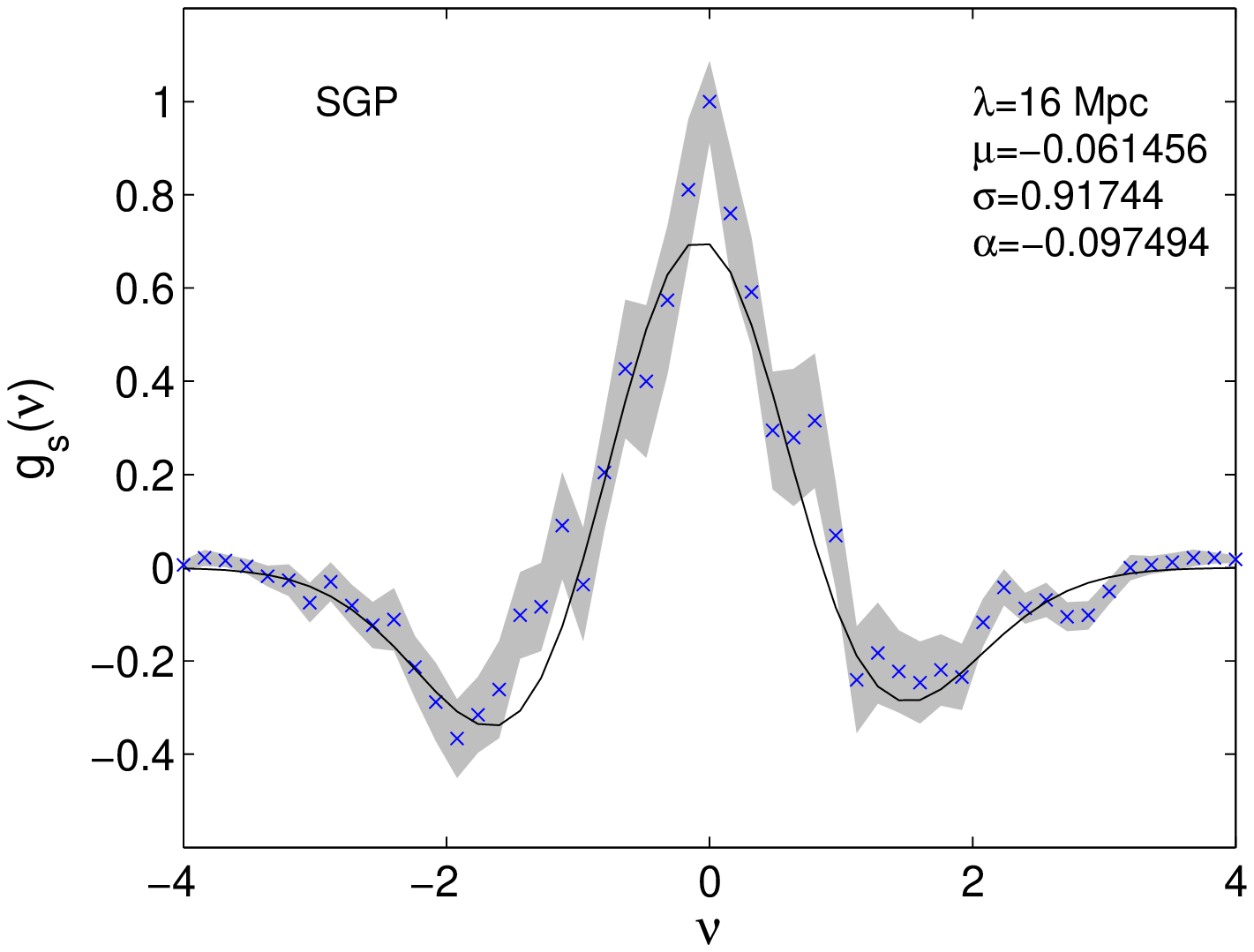}
\end{minipage}
\hspace{0.2cm}
\begin{minipage}[c]{0.48\textwidth}
\centering
\includegraphics[width=78mm]{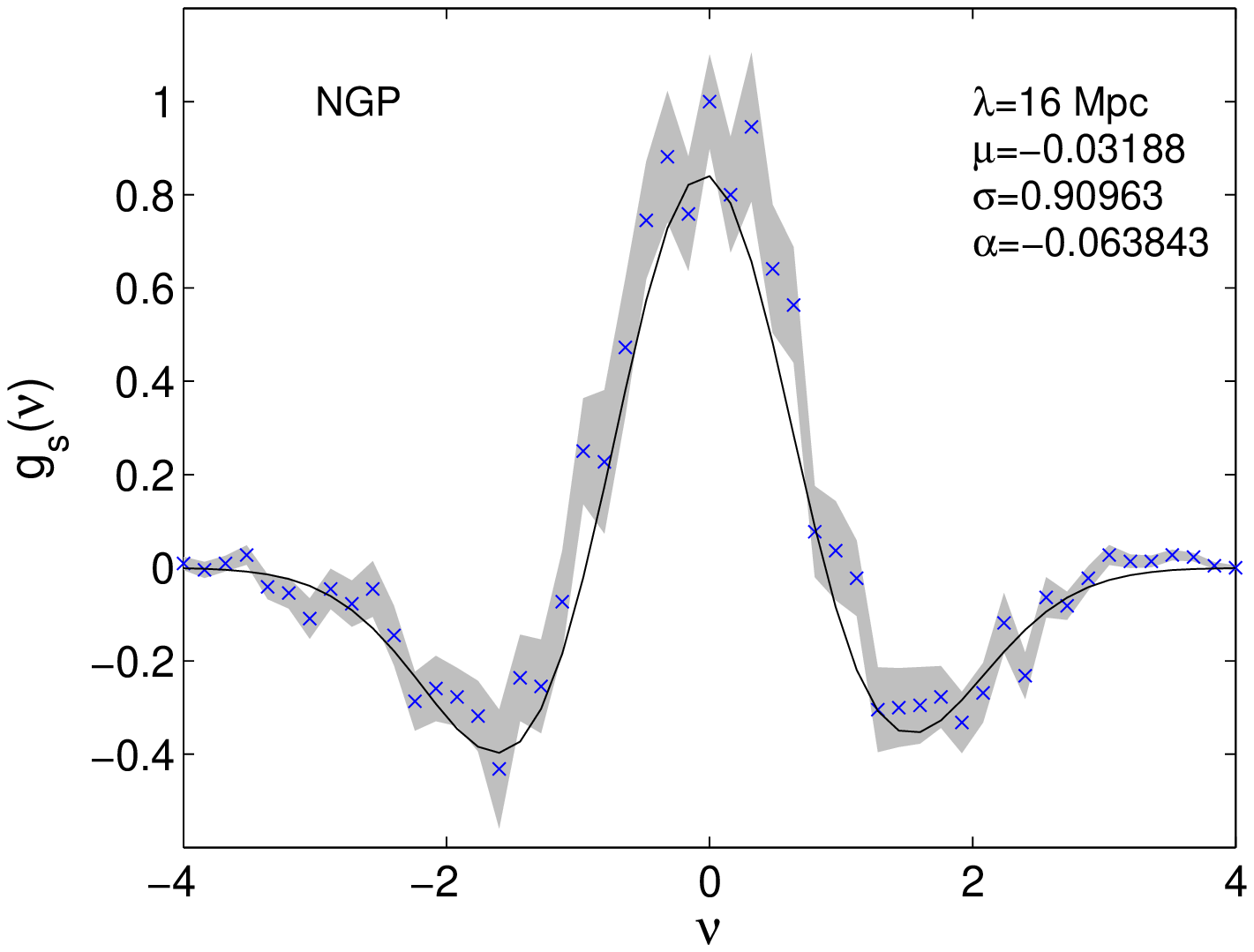}
\end{minipage}
\caption{As per Figure~\ref{fig:2df_genus}, where now the model is generated using the four-parameter transformation described in the body text. The fit is the minimum in reduced $\chi^2$ over the region $-3.5\le\nu\le3.5$.}\label{fig:2df_genus_asymm}
\end{figure*}

To determine the error bars on our sample, we have used the bootstrap resampling technique~\citep{Efron82,Barrow84}, in which we create new samples by drawing the full number of objects from the original distribution,  with replacement. The resamples are the same size as the 2dF data, but some objects are counted twice or more, and some do not appear at all (the probability of drawing exactly the data is vanishingly small for a data set of this size). The variation between resamples provides a measure of the spread in the sampling distribution of the parameter we are trying to estimate, i.e. the genus curve of the galaxy distribution.

We generated 40 resamples at smoothing lengths of 8, 10, 12, 14, 16, 18 and 20 Mpc for both the SGP and NGP slices. The standard deviation of the resamples provide error bars for each of the curves that were overlaid in Figure~\ref{fig:2df_slice}. The genus curves for 8, 12 and 16 Mpc are shown in Figure~\ref{fig:2df_genus}. 
The best fits of these models are shown in Figures~\ref{fig:2df_genus} and~\ref{fig:2df_genus_asymm}.

\subsection{Comparison with volume-limited samples}
Flux-limited samples receive the full benefit of the catalogue sampling, but require an accurately determined selection function. A retrospective comparison is possible using samples limited by volume, which remove the redshift-dependent selection of sources by omitting those too faint to be seen throughout the sample volume~\citep[a good description of the route to follow is given by][]{MartinezSaar}. 

\begin{figure*}
	\centering
	\includegraphics[width=175mm]{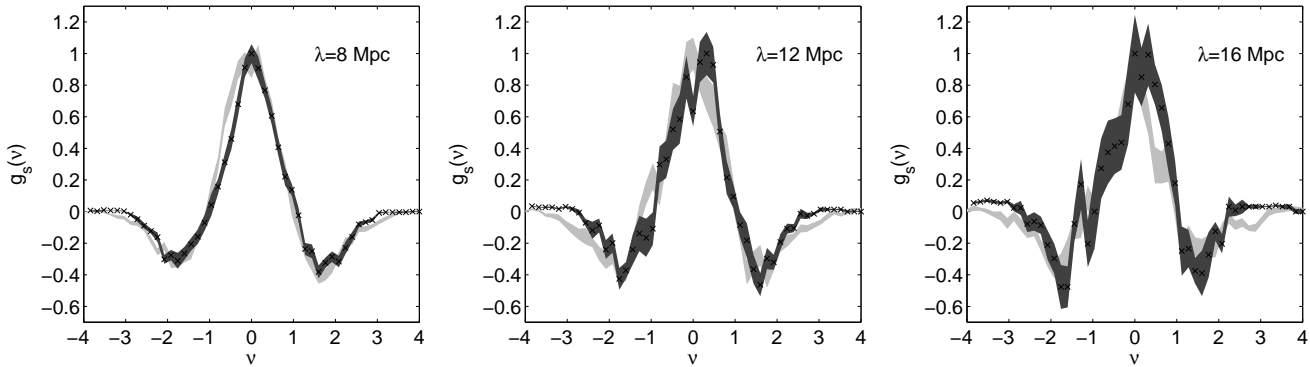}
	\caption{Genus curves for volume-limited samples from the SGP slice, shown at the fore, with crosses marking the measurement and dark shading representing $1\sigma$ intervals. The maximum redshift of the samples is $z=0.16$, selected to maximise the number of galaxies in the sample --- 49,089 of them. Underlaid in lighter shading are the corresponding curves from Figure~\ref{fig:2df_genus}.}\label{fig:comp}
\end{figure*}

The competing effects of increasing survey volume and decreasing number density create a peak in the sample size as a function of the distance limit - one is integrating progressively larger regions of space and smaller regions of the luminosity function. The choice of redshift cut that maximises the number of galaxies in the sample is $z\sim0.16$.  Figure~\ref{fig:comp} shows the resulting genus curves for the SGP slice, using the same smoothing scales as in Figure~\ref{fig:2df_genus}. The agreement with those curves is good, though the signal from the flux-limited sample is slightly better-defined; the ratio of the mean uncertainties (flux-limited:volume-limited) is 1:1.03 at 8 Mpc, 1:1.15 at 12 Mpc and 1:1.56 at 16 Mpc. Studying the progression in error bar size with scale for both sets of curves leads to the conclusion that the uncertainties are determined by (i) the number of independent subvolumes in each sample --- fixed by the total volume and the smoothing length --- as well as (ii) the precision with which the density in each subvolume is known --- determined by the sampling rate --- with the former being dominant. At 8 Mpc $\lambda$ is about 1.5 times the mean separation of points in the volume-limited sample (cf. 20 times the flux-limited sample separation), so we are just beginning to rub up against this limit. Therefore, the benefit of using flux-limited samples is that one can work with a larger volume (the choice of $z_{max} = 0.2$, while justified by Figure~\ref{fig:slopes}, is by no means exclusive), as well as probe smaller scales. 

Sparser sampling restricts the scales that can be probed, and the rapidly reducing galaxy count at redshift above $z=0.16$ removes a significant volume from consideration when using samples limited by volume. The flux-limited samples are therefore extremely successful at providing information over a broad range of scales and throughout a large region of the Universe, provided that an accurate selection function is available.

\subsection{Measurement of asymmetry}

To assess the significance of the asymmetry, we construct a composite probability distribution for the parameters of the model by calculating $\chi^2(\mathbf{P}_4)$ over a grid in four-dimensional parameter space. This can be used to determine marginal distributions for each parameter, by integrating through three of the dimensions. However, it is well worth recognising the influence of each parameter on the genus curve and constructing a physical argument to go with the model, though it is only the asymmetric parameters that provide information about the relative abundance of high- and low-density regions.

The central moments of the normal distribution are defined as
\begin{equation} M_m(x) = \int_{-\infty}^{\infty}(x-\mu)^m\frac{1}{\sqrt{2\pi}}e^{-(x-\mu)^2/2\sigma^2}dx, \label{eq:moments}\end{equation}
and there is a natural interpretation for the four parameters based around this distribution, which is the part of the analytic genus curve function that has been modified in our model. Table~\ref{table:moments} demonstrates the way in which the parameters of the model are analogous to the moments of the normal distribution. In fact, the skewness parameter $\alpha$ is not really a moment in our definition --- properly, a cubic term of the form $\alpha(x-\mu)^3$ term is required in the exponent to satisfy Equation~\eqref{eq:moments} --- but the form we have proposed represents the simplest modification available to induce an asymmetry.
\begin{table}
\centering
\begin{tabular}{|c|c|c|p{4.5cm}|}
\hline \hline
$m$ & $M_m$ & name & effect \\\hline
$0$ & $N$ & amplitude & Scales all parts of the genus curve equally \\
$1$ & $\mu$ & mean & Determines position of peak without altering the form by sliding the curve along \\
$2$ & $\sigma$ & variance & Broadens or narrows the peaks and troughs of the genus curve equally\\
$3$ & $\alpha$ & `skewness' & Asymmetrises the depths of the troughs\\
\hline \hline
\end{tabular}\caption{Table of moments of the four-parameter model, listing their natural analogues with the normal distribution moments and the effect of each of the genus curve.}\label{table:moments}
\end{table}

\begin{figure*}
\centering
\begin{minipage}[c]{0.48\textwidth}
\centering
\includegraphics[width=78mm]{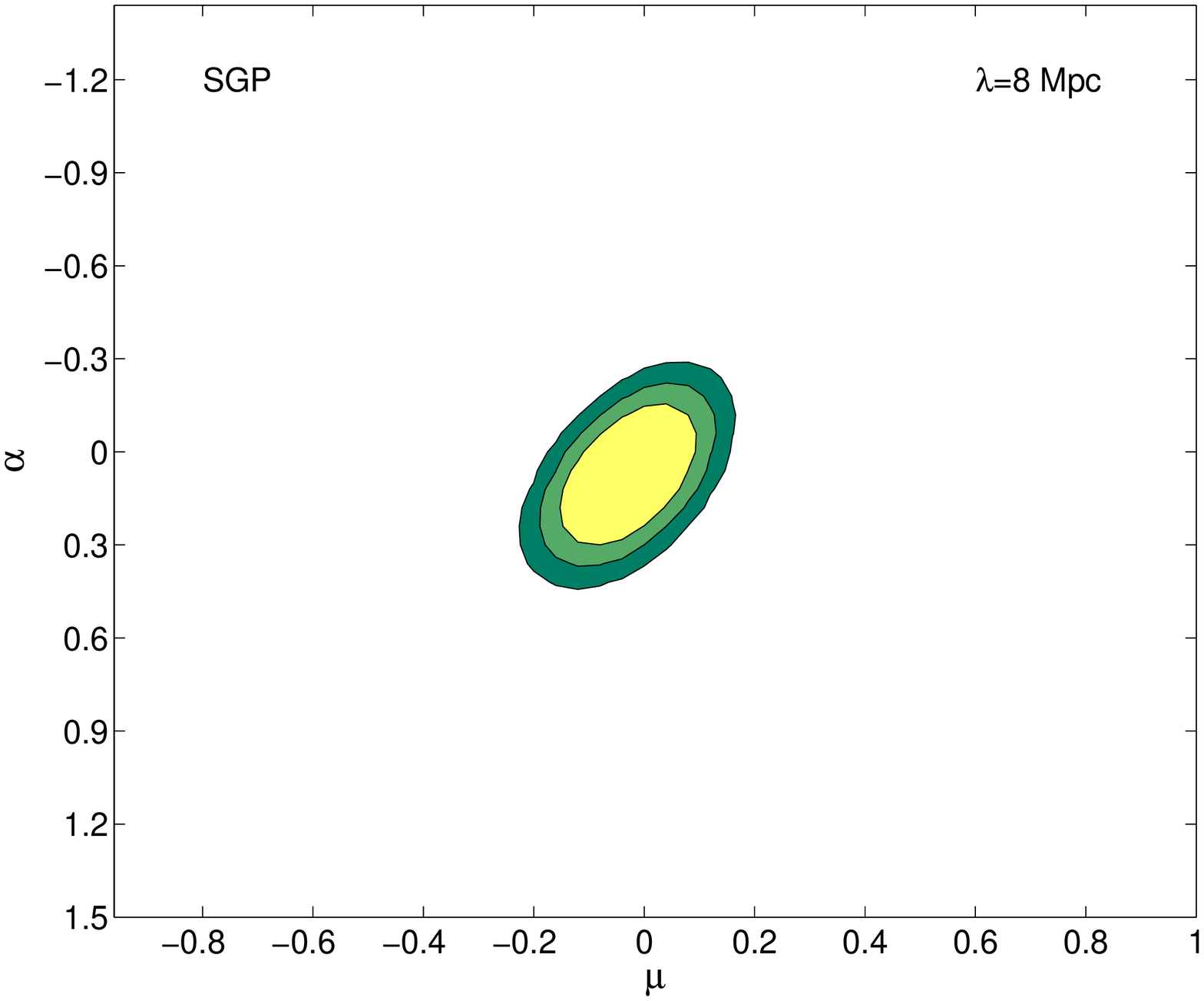}
\end{minipage}
\hspace{0.2cm}
\begin{minipage}[c]{0.48\textwidth}
\centering
\includegraphics[width=78mm]{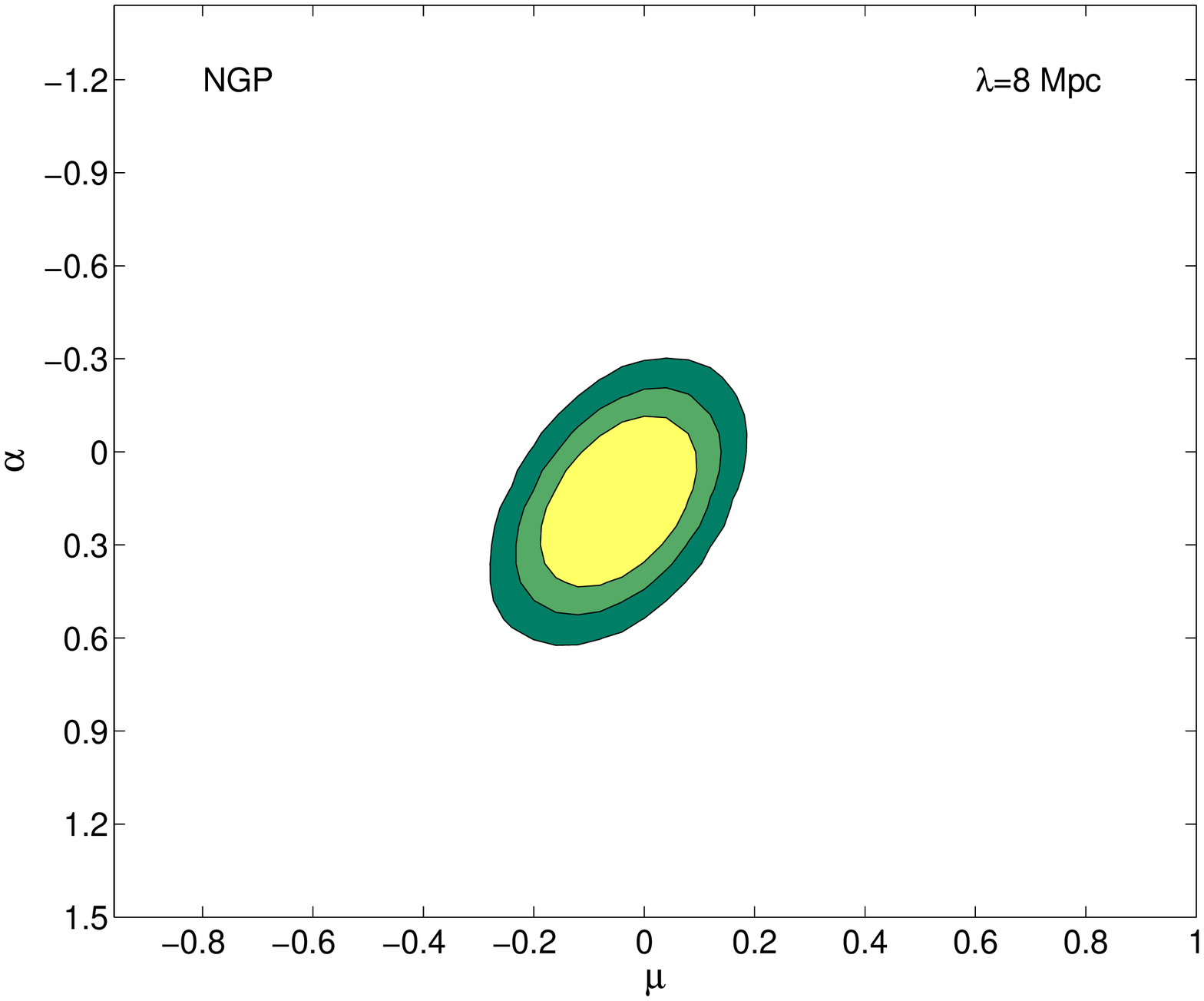}
\end{minipage}
\vspace{0.5cm}
\begin{minipage}[c]{0.48\textwidth}
\centering
\includegraphics[width=78mm]{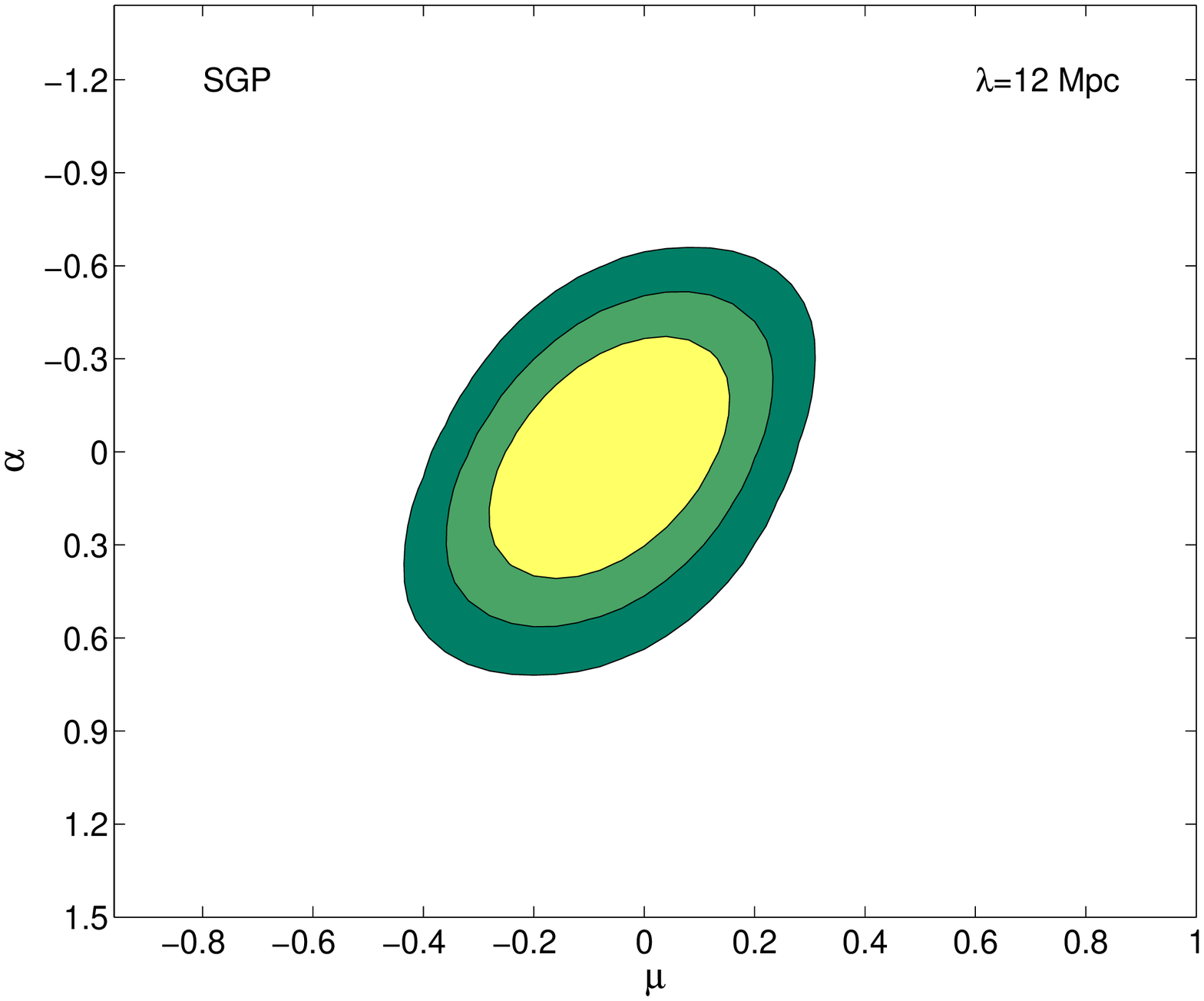}
\end{minipage}
\hspace{0.2cm}
\begin{minipage}[c]{0.48\textwidth}
\centering
\includegraphics[width=78mm]{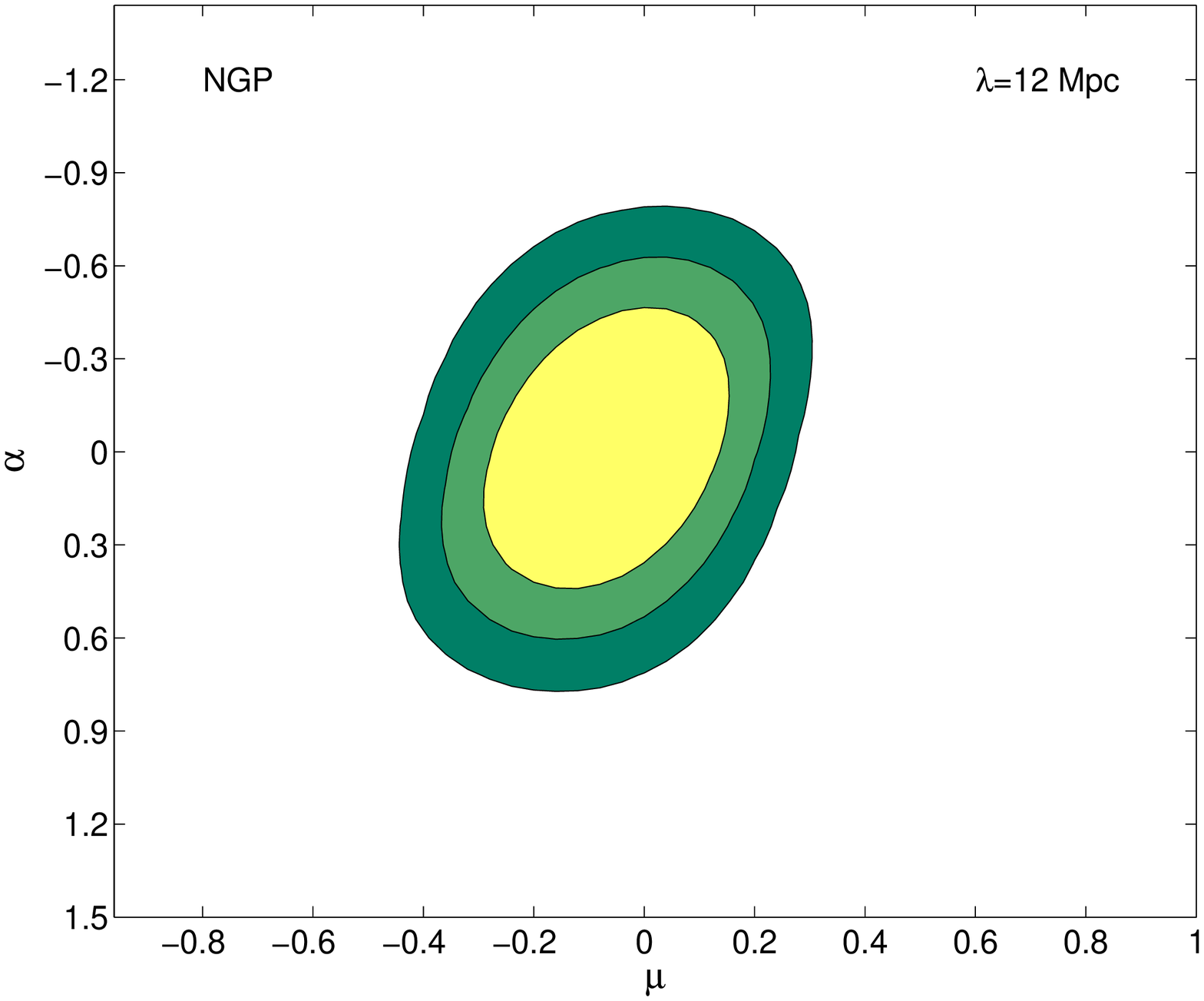}
\end{minipage}
\vspace{0.5cm}
\begin{minipage}[c]{0.48\textwidth}
\centering
\includegraphics[width=78mm]{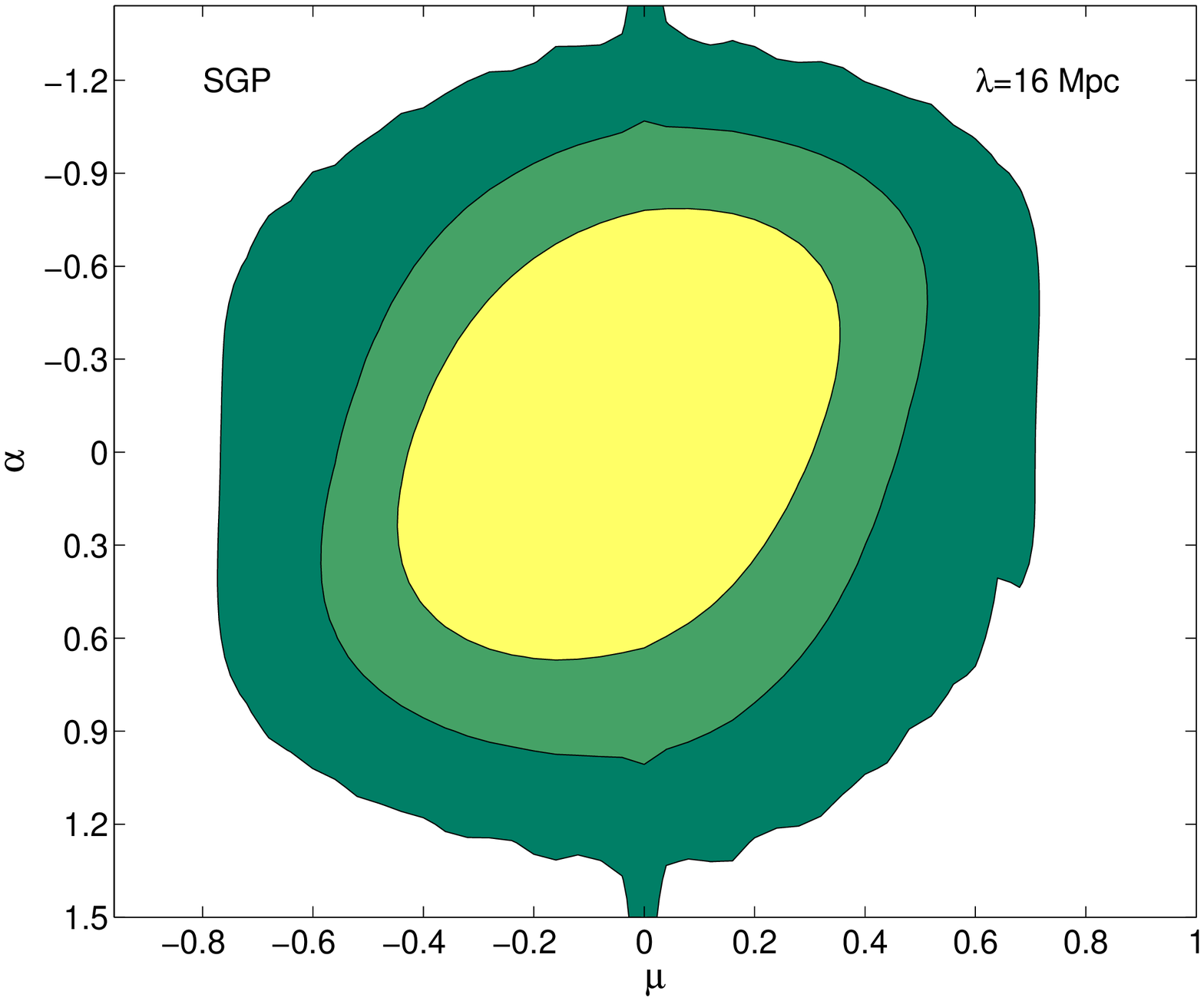}
\end{minipage}
\hspace{0.2cm}
\begin{minipage}[c]{0.48\textwidth}
\centering
\includegraphics[width=78mm]{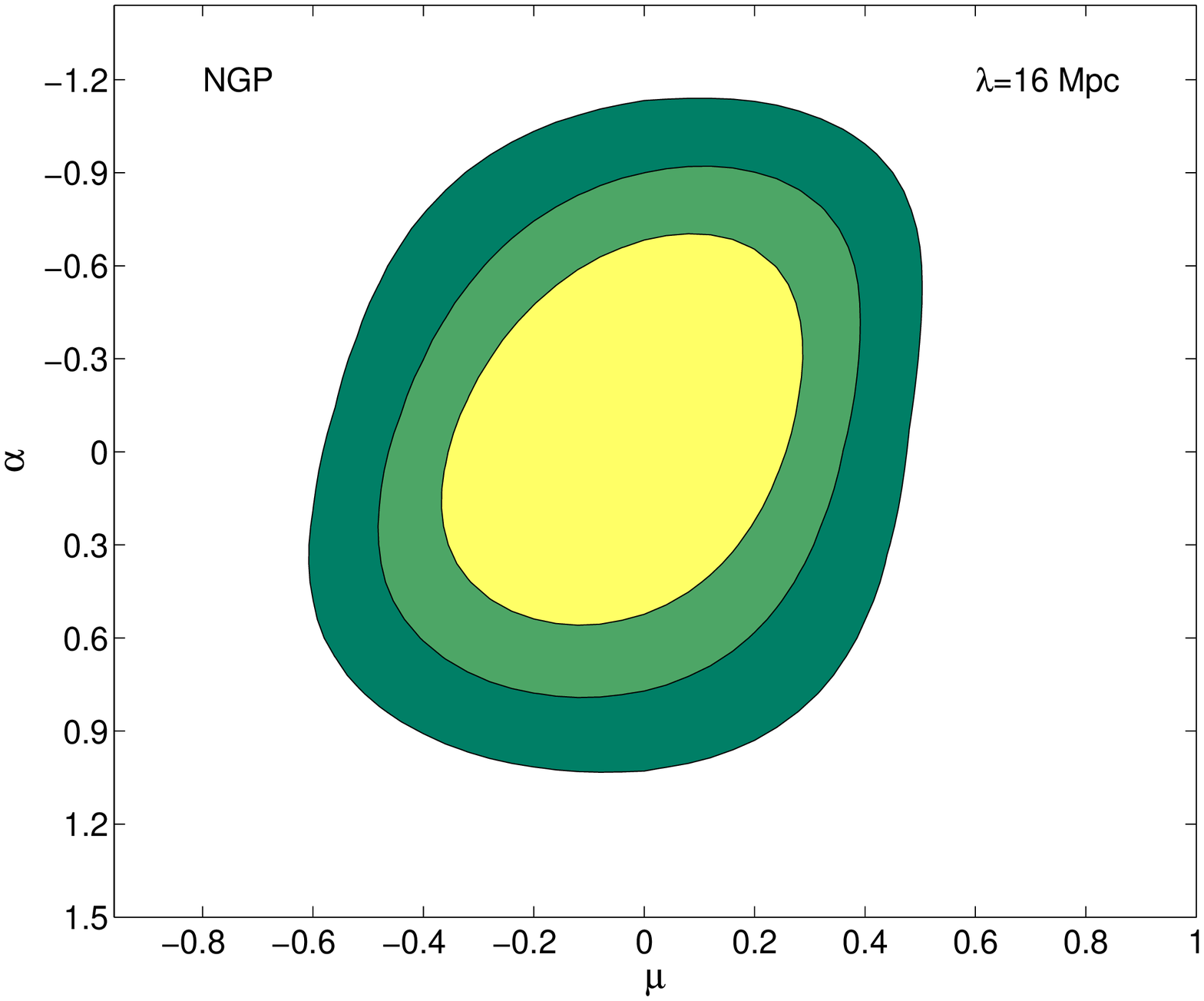}
\end{minipage}
\caption{Distribution in probability of the skewness and central tendency parameters for the SGP (left) and NGP slices on scales as in Figure~\ref{fig:2df_genus}. The curve is generated by comparing the values of $\chi^2(\mu,\alpha)$ after marginalising out the symmetric parameters in the model. On all scales, these parameters remain close to zero, consistent with the prediction of inflation, with the constraint on this result weakening due to the influx of noise on larger scales. The contours are drawn at $1\sigma$,$1.5\sigma$ and $2\sigma$.}\label{fig:2df_alpha}
\end{figure*}

As was noted above, the parameters that are most relevant here are those that invoke an asymmetry in the genus curve, i.e. $\alpha$ and $\mu$. We have produced joint probability distributions for these parameters by assuming that for given $(\alpha,\mu)$, the observed data represent independent, identically distributed realisations, so that, invoking Bayes' Theorem~\citep{Jaynes03},
\begin{equation}p(\mu,\alpha|D)\propto p(\mu,\alpha)p(D|\mu,\alpha) \propto e^{-\chi^2(\mu,\alpha)/2}\label{eq:alpha}\end{equation}
where $p(\mu,\alpha)$ represents the prior knowledge that we have about the distribution of $\mu$ and $\alpha$ --- as we have none, this is set to be uniform. In practice, we are creating the simplest intuitive probability distribution for these parameters given our observations, such that larger values of $\chi^2$ represent lower probabilities. To determine this distribution, a four-dimensional structure containing the $\chi^2$ value for points defined by $\mathbf{P}$ is created, and the symmetric parameters are marginalised out. Figure~\ref{fig:2df_alpha} shows the distribution on several scales for the model presented in Figure~\ref{fig:2df_genus_asymm}.
The value $\alpha =0$ represents a symmetric (skew-free) genus curve, and when $\mu=0$ we can interpret this value as corresponding to a Gaussian random field that has been squeezed or stretched in the wings. Deviations from $\alpha=0$ would represent skewed genus curves, not consistent with inflation, but the results here are consistent with a symmetric initial density field. The integrated probability outside the range in these figures is negligible --- from this the level contours indicating $n\sigma$ are inferred.

\begin{figure*}
\centering
\begin{minipage}[l]{0.32\textwidth}
\centering
\includegraphics[width=55mm]{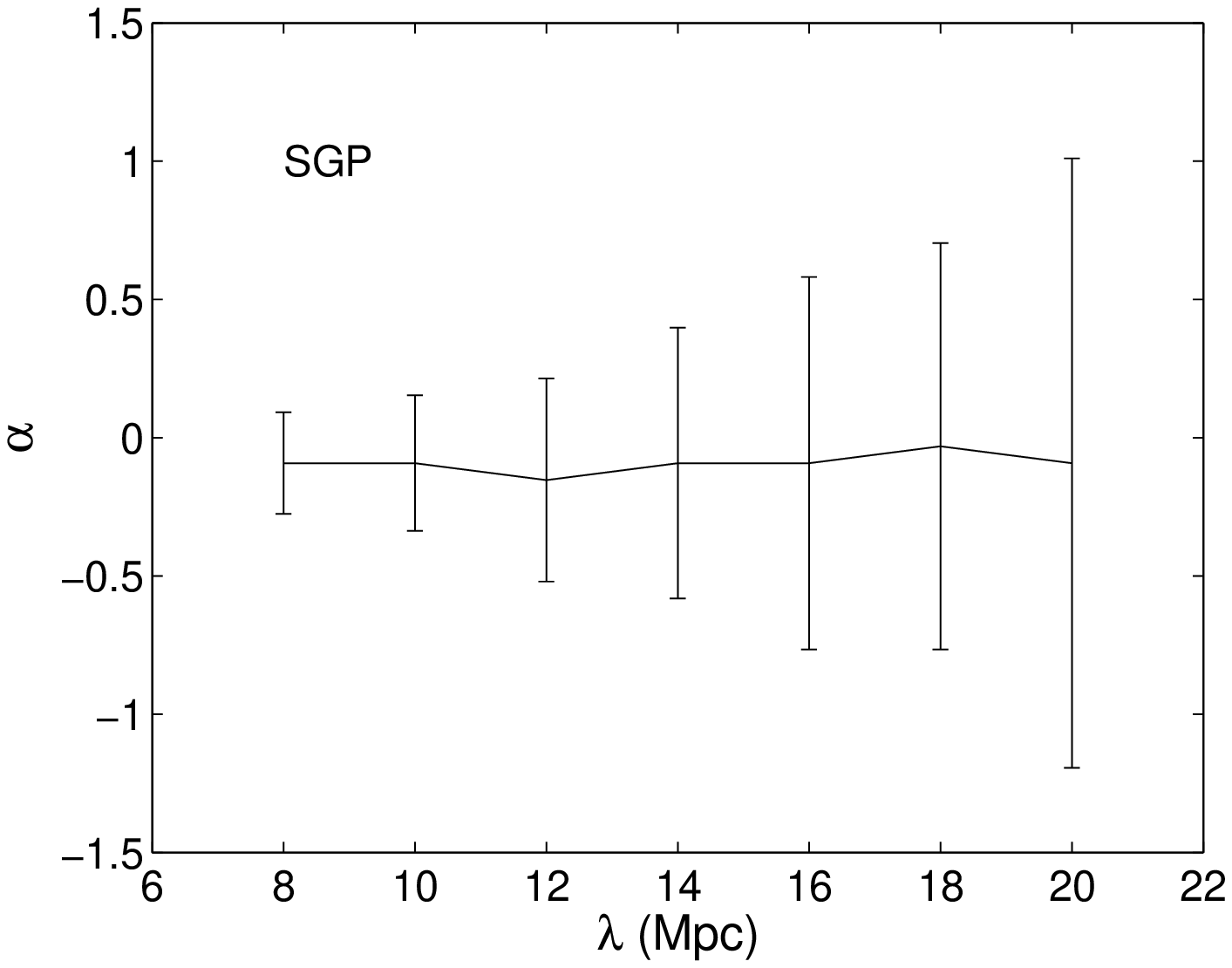}
\end{minipage}
\begin{minipage}[c]{0.32\textwidth}
\centering
\includegraphics[width=55mm]{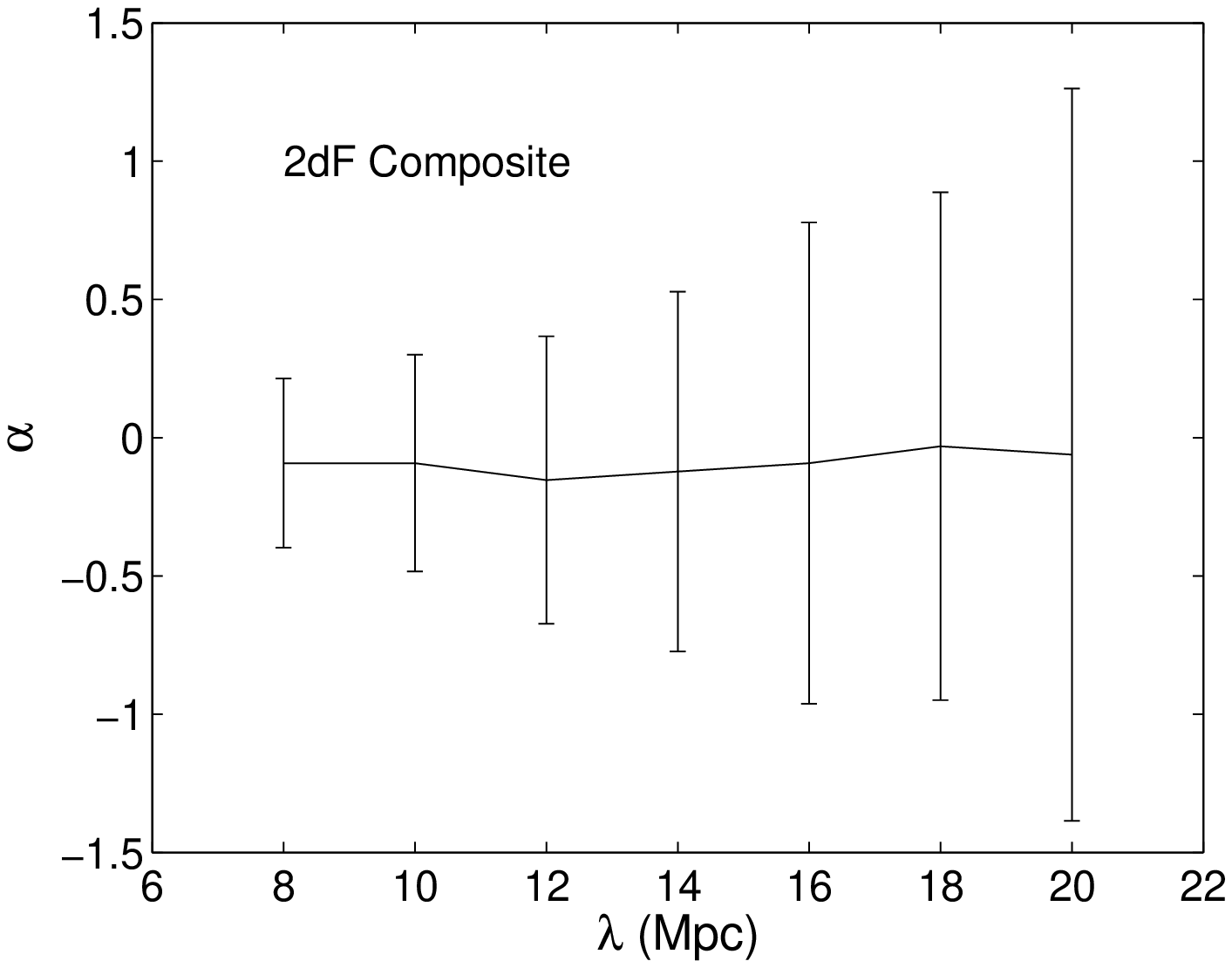}
\end{minipage}
\begin{minipage}[r]{0.32\textwidth}
\centering
\includegraphics[width=55mm]{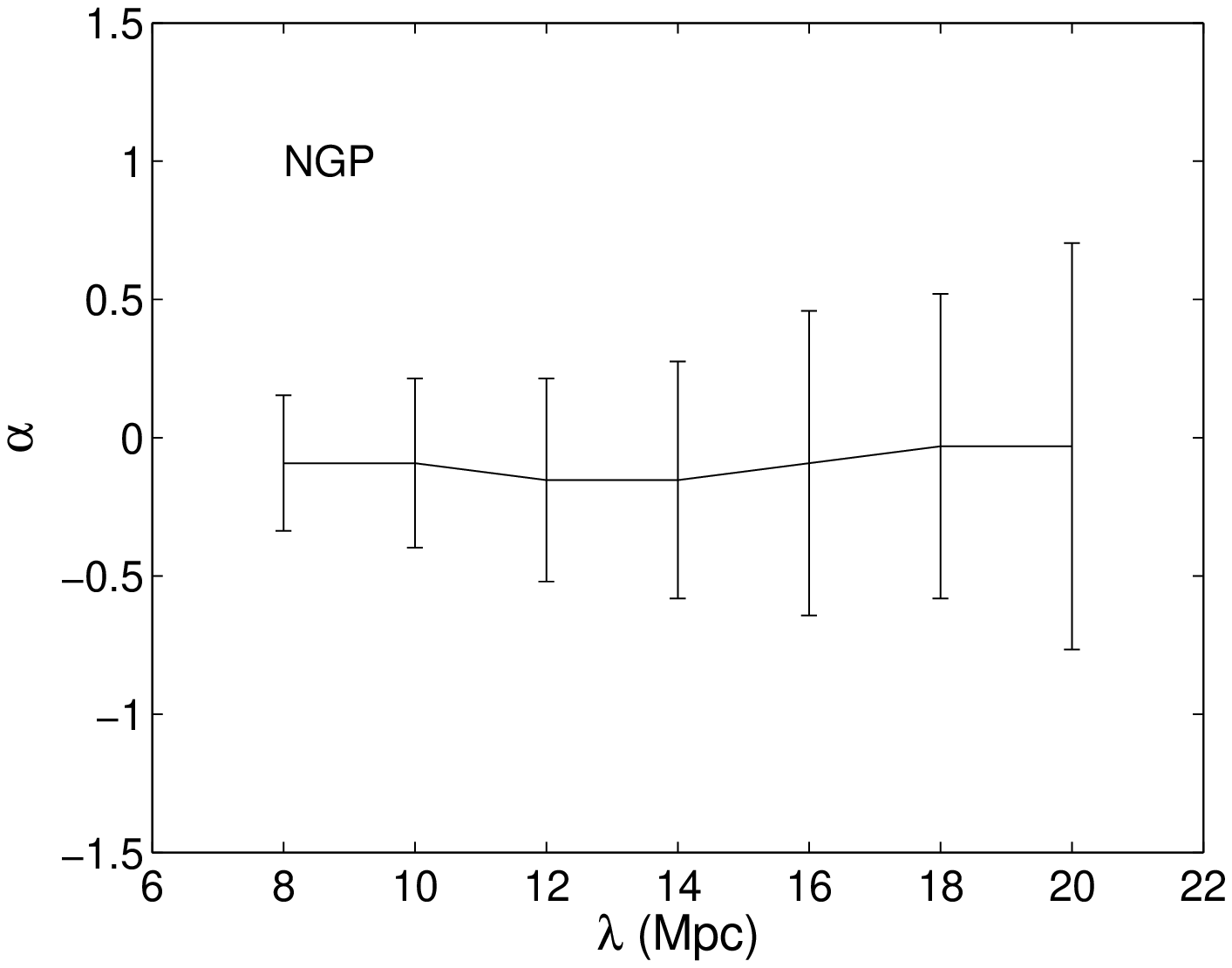}
\end{minipage}
\caption{The values of the skewness parameter $\alpha$ for the range of scales we have considered in our analysis, with 2$\sigma$ error bars, calculated from the full-width half-maximum of the distributions of $\alpha$ such as those in Figure~\ref{fig:2df_alpha}. The 2dF composite is the average of the independent NGP and SGP samples.}\label{fig:2df_final}
\end{figure*}

Figure~\ref{fig:2df_final} shows the dependence of the skewness parameter $\alpha$ on scale for both slices, and for a composite average, determined by marginalising through $\mu$. The size of the error bars suggest consistency with $\alpha=0$ on all scales observed. As the scale increases, the smoothing process reduces the number of independent points in the sample, and the width of the joint probability distribution correspondingly broadens. Consequently, the strongest constraints on the asymmetry of the genus curves occur on shorter scales (8-16 Mpc), where $|\alpha| \lesssim 1.0$ and $|\mu|\lesssim0.8$ at the $2\sigma$ level (i.e. with 95\% confidence). This indicates strongly that high- and low-density regions are symmetric at these scales. Above $\sim18$ Mpc, the constraints are weakened considerably, but remain indicative of consistency with inflation.  

\section{Conclusions}\label{sec:conclusions}
We have produced statistically complete flux-limited samples of the galaxy distribution using the 2dFGRS and its associated selection function. We have measured the topology of the smoothed distribution using the genus statistic and compared it to a Gaussian random field. These measurements are the most precise determination of the genus curve for the galaxy distribution to date. We have found the data to be consistent with Gaussian random field topology on scales of 8 Mpc and above. We interpret this as a measurement of the primordial density field, unaffected by gravitational clustering, and so find the galaxy distribution to be consistent with the prediction of inflation on scales above 8 Mpc. Moreover, we reject the presence of asymmetries in the underlying density field at the 2$\sigma$ level on scales between 8 and 16 Mpc. On these scales the topology of the galaxy distribution cannot be distinguished from that of a Gaussian random field using these methods.

Future work will focus on observations of the effects of non-linear gravitational clustering on topology. This should be apparent on scales sufficiently small to have been drawn toward a meatball topology in the time since recombination. The genus curve of the smoothed galaxy distribution on these scales should exhibit a degree of asymmetry ($|\alpha|>1$) that cannot be reconciled with an underlying Gaussian random field. The intrinsically small dispersion in genus curve measurements associated with short smoothing lengths suggests that such a result would be made with high precision indeed.

~\citet{Matsubara94} has derived an expression for the genus curve of a Gaussian random field perturbed by non-linear gravitational evolution that is valid in the weakly non-linear regime and is parametrised by $\sigma$, the rms of the density contrast in the cosmological field. However~\citet{Matsubara03} demonstrates that for a density field with spectral index of $n=-1.5$, appropriate to the galaxy distribution on below 20 Mpc~\citep{Peacock95}, the effects of the non-linearity on the genus curve are slight. At 8 Mpc, well into the non-linear regime, no significant perturbation is observed in our results, or in those of~\citet{Park05} using the Sloan data. Nonetheless, given data of sufficient precision and extending to more non-linear scales, such asymmetries may yet be seen. Understanding and quantifying this effect will be an important step in describing the behaviour of topological measurements in the non-linear regime. 

\section*{Acknowledgments}
A significant part of the research presented in this paper was conducted as a thesis for the completion of a Honours degree at the University of Sydney. This research was supported by a School of Physics Honours Scholarship, a St. Andrew's College Graduate Scholarship and the School of Physics Vacation Studentship. JBJ would like to acknowledge the support shown by members of the Institute of Astronomy at the University of Sydney, and by the astronomers at the Anglo-Australian Observatory. Peder Norberg and Shaun Cole are owed a debt of gratitude for their 2dFGRS selection function code. The authors also extend thanks to the reviewer for their helpful comments and suggestions, and Taka Matsubara for assisting us in clarifying the behaviour of genus curves in the non-linear regime.

\bibliographystyle{scemnras}
\bibliography{ms}

\appendix

\bsp

\label{lastpage}

\end{document}